\documentclass[a4paper]{JHEP3}
\usepackage{epsfig}
\usepackage{graphicx}
\usepackage{amssymb}
\usepackage{latexsym}

\newcommand{\be}{\begin{equation}}
\newcommand{\ee}{\end{equation}}
\newcommand{\br}{\begin{eqnarray}}
\newcommand{\er}{\end{eqnarray}}

\title{Effective String Theories(EST) of Yang-Mills Flux Tubes}
\author{N.D. Hari Dass\\
IMSc,Chennai,INDIA\\
Email: \email{ndhari.dass@gmail.com}}

\abstract{This chapter explains the concept of \emph{Effective String Theories}(EST), and 
their success in explaining the results that Yang-Mills flux tubes behave, to a high degree of accuracy, like Bosonic Strings(BST). 
It describes EST's of L\"uscher and Weisz, and their principal conclusions. It then discusses the Polchinski-Strominger 
EST's.  which are valid in all dimensions. It then describes the works by Drummond, and, the author and Peter Matlock, which extended the 
analysis to $R^{-3}$ order and showed, that even at that order the spectrum is that of BST.  The chapter analyses the 
issues of string momentum in higher orders. It discusses at length the powerful covariant calculus, to systematically construct EST's to 
arbitrary orders. The most general actions to $R^{-7}$ order are shown to be governed by just two parameters. The works of Aharony and 
collaborators on the spectrum of EST's, both in static and conformal gauges,to $R^{-5}$ order, and their results that even at that order 
the ground state energy remains the same as that of BST, but the excited spectrum gets corrected for $D\,>\,3$, are explained. It discusses 
the simulation results for excited states. It also discusses the AdS-CFT approaches and thickness of flux tubes.Recent works by Ambjorn,Makeenko, and Sedrakyan on the 
path-integral approaches to this issue are also discussed and compared with the other approaches. It concludes with remarks on
the significance of the results for QCD. 
}


\begin{document}
\section{Introduction}
\label{sec:ch-intro}
In chapter 21 we discussed in detail the numerical investigations of flux tubes. The investigations till around 1995 had established rather
accurately the L\"uscher term in the static $Q{\bar{Q}}$-potential:
\be
\label{eq:ch22-lterm}
V(R)\,=\,\sigma\,R\,-\,\frac{(D-2)\pi}{24\,R}\,+\ldots
\ee
The remarkable feature of the L\"uscher term is its \emph{universality}. It only depends on the space-time dimension D, and on no other
details of the gauge theory like the structure of the gauge group, the representation to which quarks belong etc. The accuracy in the 
determination of this potential was further increased by the very high accuracy studies of L\"uscher and Weisz in 2002 \cite{luscherweisz2002} 
who utilised the path-breaking \emph{Multilevel} algorithm \cite{luscherweisz2001}, which can deliver exponential reduction in statistical 
errors. L\"uscher and Weisz also analysed these results as well as the spectrum \cite{luscherweisz2002,luscherweisz2004}. 
These studies were extended further by Hari Dass and Majumdar \cite{ndhpushan2005,ndhpushan2006,ndhpushan2007,ndhpushan2008} who too made 
use of the multilevel algorithm to push investigations 
in $D=3$ SU(2) and $D=4$ SU(3) lattice gauge theories to larger $Q{\bar{Q}}$-separations than before. They found reasonable 
evidence for universality of even higher-order (in $\frac{1}{R}$) corrections to eqn.(\ref{eq:ch22-lterm}). In particular , they found that
the data in both $D=3$ and $D=4$ favour the \emph{truncated Arvis Potential}:
\be
\label{eq:ch22-truncarvis}
V_{arvis}^{trunc}(R)\,=\, \sigma\,R\,-\,\frac{(D-2)\pi}{24\,R}\,-\,\frac{(D-2)^2\pi^2}{1152\,\sigma\,R^3}\,+\ldots
\ee
Of course, both the accuracies as well as the largest $Q{\bar{Q}}$-separations probed by Hari Dass and Majumdar need to be pushed even further
to confirm that this is indeed the behaviour of the static potential.

The coefficient of the term linear in $R$, $\sigma$, is called the string-tension. In the Monte Carlo studies of the flux tube described in
the previous chapter 21, the lattice renormalization scheme adopted was the one in which this physical string tension was held fixed. The
difference between physical and lattice quantities is explained in chapter 20 on Lattice Gauge Theories, as well as in chapter 21.

A puzzling feature here, and something that bothered this author in the beginning, is that these expressions were working well even for
$D=3,4$ while their derivations \cite{arvis1983}, being based on bosonic string theory, should only be valid for $D=26$. The derivation of 
the full Arvis 
potential that yields the above when truncated to order $\frac{1}{R^3}$ as well as the $D=26$ restrictions are discussed at length in 
chapter17. The calculation of the L\"uscher term was also based on the concept of effective strings(bosonic), inspired by a parallel
found by Nambu between expectation values of large Wilson loops and certain string theoretic constructs \cite{nambustr1979}.

The main objective of this chapter is to discuss such effective string theory descriptions of these results that also explain why it is 
legitimate to use the above-mentioned results even when $D\ne\,26$. Before going into the details, it is important to clarify the concept of
an effective string theory. It is more or less in the same spirit as the \emph{Effective Field Theories} that we discussed in
chapter 18, in particular the effective field theories of strong interactions. The spirit there was to construct field theories with some
effective low energy degrees of freedom in such a way that the symmetry content of the anticipated microscopic theory (in this case the 
$SU(2)\times\,SU(2)$ symmetry of QCD) was preserved in the effective theory, but such requirements as \emph{renormalizability} or equivalently,
good high energy behaviours are no longer required. In that spirit, it could be hoped that effective string descriptions exist which retain
the symmetry content of string theories, but only restricted to such physical circumstances (to be specific, the large distance behaviour)
that issues like restrictions to $D=26$ are no longer warranted.

Indeed such a clue is provided by the full form of the Arvis potential itself (reader is urged to check details discussed in chapter 17):
\be
\label{eq:ch22-arvisfull}
V_{arvis}(R)\,=\sqrt{\sigma^2\,R^2\,-\,\frac{(D-2)\pi}{12}\,\sigma}
\ee
Below a critical $R_c\,=\,\sqrt{\frac{(D-2)\pi}{12\,\sigma}}$, the potential turns complex reflecting the \emph{tachyonic instability} of
the bosonic string theory. On the other hand, no such instability can occur in the full theory i.e. QCD which is a perfectly unitary
theory. Therefore, any effective string description can only be expected to be a reasonable description only for large R. The other notable
feature of both the L\"uscher term and the Arvis potential is the dependence on D being only through the combination $D-2$. In string theory
(see chapter 17 for an extensive discussion of this important point) this is a reflection of only the transverse vibrations being physical. That
in turn is a consequence of the world-sheet general coordinate invariance of the Nambu-Goto action. Since the Nambu-Goto action does not
involve any world-sheet metric, this is a true \emph{symmetry} unlike general coordinate invariance in General Relativity where there is
an intrinsic metric. We have alluded to this important distinction between symmetries and invariances already in chapter 18. In the case
of the Nambu-Goto action for string theory, this subtle but important distinction is not often appreciated. In the context of effective
string theories, it is this symmetry that plays a significant role (as did chiral symmetry, for example, in the effective description of strong interactions).

While we have mostly emphasized the static-potential which corresponds to the ground state of a string with both ends fixed, the Arvis
potential has an easy generalization to the excited states. We record that here for the sake of discussions later on:
\be
\label{eq:ch22-arvisexc}
E_{n,arvis}(R)\,=\,\sqrt{\{\sigma^2\,R^2\,+\,2\pi\sigma[-\frac{(D-2)}{24}\,+\,n]\}}
\ee
Again, for use in later parts of this chapter, we truncate this also to order $\frac{1}{R^3}$ and rearrange the terms to write it as
\be
\label{eq:ch22-truncarvisexc}
E_{n,arvis}^{trunc}(R)\,=\,V_{arvis}^{trunc}\,+\,\frac{\pi}{R}\,n\,+\,\frac{\pi^2}{R^3}\cdot\frac{1}{2\sigma}\,n(\frac{(D-2)}{12}-n)
\ee

The basic variables of effective string theories are also $X^\mu(\xi^a)$ as in the string theories described in chapter 17, with $\xi^a$
the world-sheet coordinates, usually denoted by $\sigma,\tau$, though with our current notation $\sigma$ for string-tension this is bound
to cause some confusion! In what follows, we first discuss effective string theories as discussed by L\"uscher and Weisz(LW) in their 2002
work, and then the effective string theories first propounded by Polchinski and Strominger(PS), and further extended by  the author, along with his 
coworkers Peter Matlock and Yashas Bharadwaj. In the parlance of string theory(again see chapter 17 for details), the L\"uscher-Weisz
effective theory is an example of the \emph{non-covariant} approach, while the Polchinski-Strominger approach is that of the \emph{covariant}
approach. The L\"uscher-Weisz choice can also be described as the \emph{static} gauge (also \emph{orthogonal} gauge). But such a nomenclature
is tied up with the symmetries of the classical action; in the string theory described in chapter 17, the classical action was the 
Nambu-Goto action and the symmetries were the world-sheet general coordinate invariances. It is only such underlying symmetries or
invariances that can guarantee the equivalence of such approaches. A case in point is gauge invariance, which is not a symmetry, but 
nevertheless guarantees the equivalence of descriptions with different choices of gauges. Another important difference is that while LW
use the path-integral formulation with a classical action as input, the PS approach is canonical (operator based).
\section{L\"uscher-Weisz Effective String Theories}
\label{sec:ch22-lweff}
The effective Bosonic string theory adopted by L\"uscher and Weisz is described first in their 2002 paper \cite{luscherweisz2002} and 
elaborated in great detail
in their 2004 paper \cite{luscherweisz2004}. Following Nambu's ideas, they express the Polyakov loop correlators as a functional integral 
for an effective string
theory governed by an action. They take this functional integral to be over all two-dimensional string world-sheets bounded by rectangular
boundaries composed of the two Polyakov loops. When $R$ and $T$ are very large (the scale is set by the string tension $\sigma$), this string 
functional integral is dominated by the so called \emph{minimal surface}, which is spanned by the two Polyakov loop(line)s as well as 
\emph{straight lines} along a spatial direction. Physically this amounts to a string configuration that is a rigid rod. The rest of the 
functional integral is evaluated by standard techniques for integrating over the fluctuations from the minimal surface whose area is just $RT$.

L\"uscher and Weisz characterize the fluctuations by only the $D-2$ transverse degrees of freedom. This is by no means obvious in a
effective string theory. Of course the L\"uscher term only depends on $D-2$ pointing to some justification for this, but it is by no
means obvious that higher corrections to the L\"uscher term also imply this. In fact, as shown by Dietz and Filk \cite{dietzfilk1983}, 
a large class of string 
actions give rise to the same L\"uscher term making it not a particularly sensitive test of effective string theories. This will become
clearer as we go along. As seen from our discussions in chapter 17, the transversality of the physical degrees of freedom is really a 
consequence of the symmetry of the action and furthermore, that action (the Nambu-Goto action) is proportional to the area of the world sheet.
That L\"uscher and Weisz state that the minimal surface dominates the functional integral also points to a classical action that is just
the area of the world sheet. But they do not explicitly state this. This is important as many of their results using the quantum spectrum
can actually be understood directly in terms of the symmetries of the classical action in so far as results to order $\frac{1}{R^3}$ are
concerned. More on this later.
\subsection{Leading Order Analysis}
\label{subsec:ch22-lweffleading}
In their 2002 paper itself LW set up the essentials of their effective string theory. They label the world-sheet coordinates by $(z_0,z_1)$,
and their $(D-2)$ transverse fields $h(z)\,=\,(0,0,h_2(z)\ldots\,h_{D-1}(z))$. The string is taken to be fixed at the two ends $z_1=0,R$.
This amounts to the transverse fields obeying Dirichlet boundary conditions at the two end-points. They take their leading action to be
of the form
\be
\label{eq:ch22-lwloact}
S_{eff}\,=\,\frac{1}{2}\,\int_0^T\,\int_0^R\,dz_0dz_1\,\partial_a\,h\cdot\,\partial_a\,h\,+\ldots
\ee
where $\cdot$ denotes the `scalar product' in transverse space. They say this form is dictated by symmetry considerations but do not specify what those are. The
author guesses they have the two-dimensional Lorentz invariance in mind. Accordingly they say the higher order terms are Lorentz-invariant
combinations of derivatives of $h$. In their treatment, $h$ is dimensionless reflecting the fact that the actual fluctuations have been scaled
by $\sqrt{\sigma}$. This is also reflected in the dimensionless coefficient $\frac{1}{2}$ for the leading term. Of course, the higher order
terms will necessarily have dimensionful coefficients rendering the theory \emph{non-renormalizable}. We shall have more to say on this
as we go along.

Lastly, they assume periodic boundary conditions along the $z_0$-direction, i.e. $h_i(T,z_1)\,=\,h_i(0,z_1)$. The leading order string 
functional integral can be evaluated exactly and has been known for a long time in string-literature (see their paper for a list of sources;
also Polchinski's vol I on String Theory \cite{polchinskistrI}). 
This then leads to their exact result
at this order
\be
\label{eq:ch22-lwloexact}
\langle\,P(x)^*P(y)\rangle\,=\,e^{-\sigma\,RT\,-\,\mu\,T}[det(-\Delta)]^{-\frac{(D-2)}{2}}
\ee
where $\Delta$ is the Laplacian on a two-dimensional cylinder of height $R$ and circumference $2\pi\,T$ with Dirichlet boundary conditions.
The quantity $\mu$ is a free parameter arising essentially out of renormalization effects. It already made its appearance in the results
of Ambjorn et al \cite{ambjorn84}. Fortunately, it only contributes an additive constant to the static potential and is hence devoid of 
any physical
significance. As LW point out, such terms arise in the gauge theory itself arising out of the need to renormalize Wilson loop
expectation values as well as Polyakov loop correlators. It is however curious that neither the Arvis quantization of fixed-end strings with
the Nambu-Goto action, nor the Polyakov-Strominger(PS) effective string description presented in the next section, require such explicit
renormalizations.
 
The result for the determinant is $det(-\Delta)\,=\,\eta(q)^2$ where $\eta(q)$ is the Dedekind-$\eta$ function, given explicitly by
\be
\label{eq:ch22-dedeta}
\eta(q)\,=\,q^{\frac{1}{24}}\,\prod_{n=1}^\infty\,(1\,-\,q^n)\quad\quad\,q\,=\,e^{-\pi\,\frac{T}{R}}
\ee
For large $T$, one can expand in powers of $q$ which straight away yields the spectral decomposition for the Polyakov loop correlators! The
result for the energy levels is
\be
\label{eq:ch22-lwlospect}
E_n(R)\,=\,\sigma\,R\,+\,\mu\,+\,\frac{\pi}{R}\{-\frac{(D-2)}{24}\,+\,n\}
\ee
The first few degeneracies are given by $w_0=1,w_1=(D-2),w_2=\frac{1}{2}\,(D-2)(D+1)$(this should be compared with results from the
oscillator representation of dual resonance models worked out as an example in chapter 16). Comparison with eqn.(\ref{eq:ch22-lterm})
shows complete agreement for the static-potential to this order as well as the uniform  spacing of $\frac{\pi}{R}$ of the excited levels.
\subsection{A Possible Boundary Term}
\label{subsec:ch22-lwboundary}
Higher order corrections to $S_{eff}$ will typically involve coefficients with negative-integer mass dimensions. A mass dimension $-1$
term is only possible as a boundary term, which can certainly be there for the case of open-strings, while there can be no such terms
for closed strings. One such boundary term, the simplest, considered by LW is
\be
\label{eq:ch22-lwbound}
S_1\,=\,\frac{b}{4}\,\int_0^T\,dz_0\,\{(\partial_1\,h\cdot\partial_1\,h)|_{z_1=0}\,+\,(\partial_1\,h\cdot\partial_1\,h)|_{z_1=R}\}
\ee
with b, a free parameter, having mass dimension -1. Power counting arguments would say this is a non-renormalizable interaction. But as
LW point out in \cite{luscherweisz2004}, there are no ultraviolet divergences. Presumably this is because this interaction is still quadratic.
LW find corrections to the static-potential and energy level spacings to be
\be
\label{eq:ch22-lwboundcorr}
\Delta\,V(R)\,=\,-\frac{\pi(D-2)\,b}{24\,R^2}\quad\quad\,\Delta\,E\,=\,\frac{\pi}{R}\,(1\,+\,\frac{b}{R})
\ee
Clearly such a term is not present in eqn.(\ref{eq:ch22-lterm}) as the quantization leading to this term had not included such boundary
terms. Though for very large $R$ this term is sub-leading to the L\"uscher term, with accuracies that are able to probe even $\frac{1}{R^3}$
corrections, this would certainly show up unless b itself is very small. LW initially tried to account for their observed discrepancy of
about 15\% in the L\"uscher term through such a boundary term with $b=0$ in $D=3$ and $b=0.04fm$ in $D=4$. Hari Dass and Pushan Majumdar,
however found better fits without such boundary terms, and with only the universal $\frac{1}{R^3}$ terms of the truncated Arvis potential.
In \cite{luscherweisz2004}(section \S\,2.3) it is further pointed that at this order, the sole effect of this boundary interaction is 
the shift $R\rightarrow\,R-b$. In the $D=4$ studies of \cite{ndhpushan2005,ndhpushan2006}, the lattice spacing was $\frac{1}{6}$fm and 
a $b$ of 0.04 fm would amount to a noticeable shift of one quarter of lattice spacing.
In the next section we discuss the arguments of LW based on what they called open-closed string duality to argue that $b$ should actually be
zero.
\subsection{Dimension -2 corrections}
\label{subsec:ch22-dim2corr}
Motivated by several issues raised by their results in \cite{luscherweisz2002}, LW recognized the need to study even higher order corrections
to the spectrum of effective string theories. In their 2004 work \cite{luscherweisz2004} they undertook  systematic studies of the same,
and provided powerful calculational tools for the same. In particular, they focussed on potential (pun intended) corrections at $\frac{1}{R^3}$
level.

They argued that such and higher order corrections ought to be captured by interaction terms that are localized either on the world-sheet
or on its boundary. Furthermore, they emphasized, in accordance with what is known in QFT's, that terms that are removable by field
redfinitions can be dropped. We shall, however, have some comments later on about the locality aspects. Then they write down two possible
interaction terms, both dimension 2, that are in addition to the dimension 1 boundary term already discussed. These are
\br
\label{eq:ch22-lwdim2}
S_2\,&=&\,\frac{c_2}{4}\,\int\,d^2z\,(\partial_a\,h\cdot\partial_a\,h)(\partial_b\,h\cdot\partial_b\,h)\nonumber\\
S_3\,&=&\,\frac{c_3}{4}\,\int\,d^2z\,(\partial_a\,h\cdot\partial_b\,h)(\partial_a\,h\cdot\partial_b\,h)
\er
At this point LW introduce the notion of \emph{Open-closed string duality} and draw many conclusions from it. In the present context, two
of the important conclusions are i) vanishing of the boundary term i.e. $b=0$ in all dimensions, and, ii) a linear constraint between
$c_2$ and $c_3$. We discuss these aspects in the next subsection.
\subsection{Open-closed String Duality}
\label{subsec:ch22-stringduality}
LW point out two equivalent interpretations of the Polyakov loop correlators. First is by considering the Polyakov lines to run along
the time axis (with periodic boundary conditions in the time direction); this leads to the gauge theory interpretation of their
correlation function as propagation in the presence of a $Q{\bar{Q}}$-pair, and the subsequent effective string interpretation in terms of
open strings. Equally well, the compact dimension with periodic boundary conditions could be taken along what was earlier thought to be 
a spatial direction. In this case, the effective string description would be that of a closed string propagation. This dual interpretation
leads to relations between the partition functions in the two pictures.

In particular, for free string propagation (by this LW mean partition functions with only the leading $S_0$ term) for example, this leads
to the relation
\be
\label{eq:ch22-freeduality}
{\cal Z}_0\,=\,(\frac{T}{2R})^{\frac{(D-2)}{2}}\,{\cal Z}_0|_{T\leftrightarrow\,2R}
\ee
To demonstrate this they make use of the elegant modular transformation properties of the Dedekind-$\eta$ function:
\be
\label{eq:ch22-etamodular}
\eta(q)\,=\,(\frac{2R}{T})^{\frac{1}{2}}\,\eta({\tilde{q}})\quad\quad\,{\tilde{q}}\,=\,e^{-\frac{4\pi\,R}{T}}
\ee
Let us first consider their analysis of the implications of this duality at the next higher order i.e. including the boundary term $S_1$.
Because of the quadratic nature of this correction, LW found a simple relation relating the partition function ${\cal Z}$ to this order
and ${\cal Z}_0$:
\be
\label{eq:ch22-boundpart}
{\cal Z}\,=\,(1\,-\,\sigma\,T\,b){\cal Z}_0\,-\,b\,\frac{\partial}{\partial\,R}\,{\cal Z}_0
\ee
i.e. the duality constraints on ${\cal Z}$ from those of ${\cal Z}_0$. One then expands the partition function for large $T$, and the dual for
large R. Upon doing so, LW found that a non-vanishing b is inconsistent with duality and they drew the first important conclusion that
$b=0$.

Because of the absence of $S_1$, standard path integral perturbation methods yield ${\cal Z}$ when the dimension 2 corrections $S_2,S_3$
are taken into account, with the result
\be
\label{eq:ch22-lwdim2part}
{\cal Z}\,=\,{\cal Z}_0\,\{1\,-\,\langle\,S_2\rangle_0\,-\,\langle\,S_3\rangle_0\}
\ee
where $\langle\ldots\rangle_0$ denote expectation values in the free theory. The required expectation values can again be expressed 
in terms of $q$, $\eta(q)$ and its derivatives \cite{luscherweisz2004,dietzfilk1983}. We skip the details and go straight to the
constraints imposed by duality for $c_2,c_3$:
\be
\label{eq:ch22-lwdim2duality}
(D-2)c_2\,+\,c_3\,=\,\frac{D-4}{2\sigma}
\ee
Before going into the consequences of this, we mention the demonstration by LW (in appendix C of \cite{luscherweisz2004}) that the full
Arvis spectrum mentioned before satisfies the open-closed string duality exactly. Consequently, the truncated Arvis potential of
eqn.(\ref{eq:ch22-truncarvis}) and energy levels of eqn.(\ref{eq:ch22-truncarvisexc}) 
also satisfy the duality constraints.

To appreciate their implications, let us look at the $\frac{1}{R^3}$ corrections that were worked out by LW in their eqns (4.9,4.10):
\br
\label{eq:ch22-lwdim2spect}
\Delta\,V(R)\,&=&\,(\frac{\pi}{24})^2\,\frac{(D-2)}{2\,R^3}\,\{2\,c_2\,+\,(D-1)\,c_3\}\nonumber\\
\Delta\,E_1\,&=&\,\frac{\pi^2}{24\,R^3}\{(12\,D-14)c_2\,+\,(5\,D\,+\,7)c_3\}
\er
Here only the results for the first excited state has been displayed, but it is more transparent to use their eqn.(6.1) for $n\le\,3$:
\be
\label{eq:ch22-lwdim2spect2}
\Delta\,E_{n,i}\,=\,\frac{\pi^2}{R^3}\,\{n[\frac{(D-2)}{12}-n]c_2\,+\,\nu_{n,i}(c_3\,+\,2\,c_2)\}
\ee
where n refers to the level and i to its splittings. LW tabulate the coefficients $\nu_{n,i}$ in their Table 2, but we shall not be
needing them.

At this point LW draw many conclusions from the duality constraints. An obvious one, according to them, is $c_3\,=\,-2c_2$ in $D=4$. We shall
shortly argue that this is so for a large class of actions in \emph{all} D. Eqn.(\ref{eq:ch22-lwdim2spect2}) also reveals that when this
condition is met, the terms with $\nu_{n,i}$ drop out completely. LW also point out that in $D=3$ the integrands of the two actions are 
the same. Hence the spectrum is only sensitive to the combination $c_2\,+\,c_3$. The duality constraint in this case reads 
$c_2\,+\,c_3\,=\,-\frac{1}{2\sigma}$. So they state that without loss of generality one can choose $c_3=\,-2c_2$. This would then yield
$c_2\,=\,\frac{1}{2\sigma}$ and $c_3\,=\,-\frac{1}{\sigma}$.

While the author agrees with these values for other reasons, he has reservations about their arguments. Firstly, if the spectrum did depend
on $c_2+c_3$ only, whose value is completely fixed by the duality constraint in $D=3$ to be $-\frac{1}{2\sigma}$, nothing more needs to be
done i.e. making the choice $c_3\,=\,-2c_2$ is unwarranted. However, even in $D=3$, 
eqn.(\ref{eq:ch22-lwdim2spect2}) does not involve $c_2,c_3$ only in the combination $c_2+c_3$, though it is true for their eqn(4.9) for $D=3$. 
This can be seen explicitly by comparing the coefficients of $c_2$ and $c_3$; 
\be
\label{eq:ch22-lw3dprob}
c_2: n[\frac{(D-2)}{12}\,-\,n]\,+\,2\nu_{n,i}\quad\quad\,c_3:\nu_{n,i}
\ee
Their equality, for $D=3$, is only possible if
\be
\label{eq:ch22-lw3dprob2}
\nu_{n,i}\,=\,n(n\,-\,\frac{1}{12})
\ee
That is, $\nu_{n,i}$ should be \emph{independent} of i. 
Their Table 2 seems to be at variance with this.

Irrespective of the actual values of $\nu_{n,i}$, as per their eqn(6.1), for given $c_2$, there is a clear dependence on $c_3+2c_2$ and
they can not set $c_3+2c_2\,=\,0$ (without loss, in their words). Nevertheless, let us accept this and see what obtains. Firstly, the 
$\nu_{n,i}$-dependence disappears. In $D=3$, the duality constraint $c_2+c_3\,=\,-\frac{1}{2\sigma}$ uniquely fixes $c_2\,=\,\frac{1}{2\sigma}$
and $c_3\,=\,-\frac{1}{\sigma}$. Consequently, the $D=3$ spectrum to this order coincides with the truncated Arvis spectrum, in agreement
with the results of \cite{ndhpushan2007,ndhpushan2008} for the static potential.

Coming to $D=4$, the duality constraint itself gives $c_3+2c_2\,=\,0$ and all dependence on $\nu_{n,i}$ vanishes, but either $c_2$ or
$c_3$ remains undetermined. But \cite{ndhpushan2005,ndhpushan2006} found agreement with truncated Arvis potential (they did not study the
excited states; we will comment on the excited states later), with no room for any free parameters. The resolution of this will be discussed
in the next section on the basis of the Polchinski-Strominger effective string theories. In that section, the issue of the legitimacy of using
relations based on string theory like the Arvis potential, which should only be valid for $D=26$, for arbitrary D values, will also be
clarified. Before that, the author wishes to discuss a purely classical perspective on $c_2,c_3$ and the relationship between them.
\subsection{Purely Classical Analysis}
\label{eq:ch22-lwclassical}
As $c_2,c_3$ are aspects of a classical action, one ought to be able to find any constraints on them in a purely classical manner without going into
the details of the quantum spectrum, as LW did. Of course, in the end there has to be consistency between the classical and quantum
approaches. Before delving into that, let us gain another perspective on the duality constraint between $c_2$ and $c_3$, and for that let
us consider those classical actions for which $c_2$ and $c_3$ do not explicitly depend on D. Then the duality constraint 
$(D-2)c_2+c_3\,=\,\frac{D-4}{2\sigma}$ can only be satisfied for all D if and if only $c_3+2c_2\,=\,0$. This according to the author, is
the primary significance of this relation though LW have discussed other interesting consequences of this relation. The immediate consequence 
of satisfying this is that $c_2\,=\,\frac{1}{2\sigma}$ and $c_3\,=\,-\frac{1}{\sigma}$, precisely the values that lead to the truncated Arvis
spectrum. In what follows, all our remarks are for order $\frac{1}{R^3}$ only.

Are there such classical actions for which $c_2,c_3$ have these properties? We first examine the Nambu-Goto action itself as we have
already argued that the dominant action for the LW analysis must be this, even though they did not explicitly state it. As discussed at
length in chapter 17, the Nambu-Goto action(we go back to the notations of chapter 17) is
\be
\label{eq:ch22-nambugoto}
S_{NG}\,=\,-{\cal T}\,\,\int\,d\tau\,\int_0^\pi\,d\sigma\,\sqrt{det(-g)}\quad\quad\,g_{\alpha\beta}\,=\,\partial_\alpha\,X^\mu\,\partial_\beta\,X^\nu\,\eta_{\mu\nu}
\ee
with ${\cal T}$ being the string tension ($\sigma$ in LW notation as also in \cite{ndhpushan2006}, for example).
Written out explicitly(see chapter 17) \cite{ggrtqstr1973,arvis1983}
\be
\label{eq:ch22-nambugoto2}
S_{NG}\,=\,-{\cal T}\,\int\,d\tau\int_0^\pi\,d\sigma\,\sqrt{(\frac{\partial\,X}{\partial\sigma}\cdot\frac{\partial\,X}{\partial\,\tau})^2\,-\,
(\frac{\partial\,X}{\partial\,\sigma})^2(\frac{\partial\,X}{\partial\,\tau})^2}
\ee
The most important aspects of the Nambu-Goto action for our purposes are the twin invariances of i)\emph{Target Space}-Lorentz invariance,
and, ii) world-sheet general coordinate invariance (sometimes fancily called diffeomorphism invariance). Both of them play crucial roles in
what follows.

The general coordinate invariance allows the use of the \emph{Static Gauge}: $X^0\,=\,\tau$. In this gauge, the straight string configuration
of a string fixed at the two end-points $\sigma\,=\,0,\pi$ is given by $X^1\,=\,\frac{R}{\pi}\,\sigma,X^i\,=\,0$. The fluctuations about this 
are given by $X^i\,=\,h^i$ with the $D-2$ transverse variables $h^i$ obeying Dirichlet-boundary conditions. Thus the $h^i$ can be
identified with LW variables. The minimal surface is rectangular with extent $T$ along $X^0$ direction and $R$ along with $X^1$-direction with
area RT, contributing ${\cal T}\,RT$ to the action.

The $(-g)$ factor is easily worked out for the configuration with fluctuations. The result is
\be
\label{eq:ch22-NGdet}
det(-g)\,=\,(\partial_\sigma\,h^i\partial_\tau\,h^i)^2\,-\,(-1\,+\,(\partial_\tau\,h^i)^2)(\frac{R^2}{\pi^2}\,+\,(\partial_\sigma\,h^i)^2)
\ee
Let us collect terms upto quadratic order in fluctuations to fix various scalings and normalizations:
\be
\label{eq:ch22-NGdetquad}
det(-g)_2\,=\,\frac{R^2}{\pi^2}\,+\,(\partial_\sigma\,h^i)^2\,-\,(\partial_\tau\,h^i)^2\cdot\frac{R^2}{\pi^2}
\ee
This immediately suggests the rescaling $\sigma\rightarrow\,\sigma^\prime\,=\,\frac{R}{\pi}\,\sigma$ so that the range of $\sigma^\prime$ is
$(0,R)$ as in the LW conventions. After this scaling the quadratic part of the action becomes
\be
\label{ch22-NGquad}
S_0\,=\,\frac{{\cal T}}{2}\,\int\,d\tau\,\int_0^R\,d\sigma^\prime\,(\partial_a\,h)(\partial_a\,h)
\ee
So a rescaling of the $h$ field by a $\sqrt{{\cal T}}$ makes the $h$-field dimensionless and yields precisely the $S_0$ of LW . Now we turn to the 
next higher order correction where the scalings done above are already incorporated. To avoid clutter, the $\prime$ on the rescaled $\sigma$
will not be shown explicitly:
\be
\label{eq:ch22-NGdim2}
{\cal L}_2\,=\,\frac{1}{2{\cal T}}\,\{(\partial_\sigma\,h^i\partial_\tau\,h^i)^2\,-\,(\partial_\tau\,h^i)^2(\partial_\sigma\,h^j)^2\}
\,-\,\frac{1}{8{\cal T}}\{(\partial_\sigma\,h^i)^2\,-\,(\partial_\tau\,h^i)^2\}^2
\ee
The first term, after some algebra, can be reexpressed as
\be
\label{eq:ch22-NGdim2p}
-\frac{1}{2{\cal T}}\,\frac{1}{2}\,\{(\partial_a\,h\partial_b\,h)(\partial_a\,h\partial_b\,h)\,-\,(\partial_a\,h\partial_a\,h)(\partial_b\,h\partial_b\,h)\}
\ee
Combining all terms, one finally arrives at
\be
\label{eq:ch22-NGdim2fin}
S_2+S_3\,=\,\int\,d\tau\,\int_0^R\,d\sigma\{\frac{c_2}{4}(\partial_a\,h\partial_a\,h)(\partial_b\,h\partial_b\,h)\,+\,\frac{c_3}{4}\,(\partial_a\,h\partial_b\,h)(\partial_a\,h\partial_b\,h)\}
\ee
with $c_2\,=\,\frac{1}{2{\cal T}}$ and $c_3\,=\,-\frac{1}{{\cal T}}$. This is precisely the form of the dim 2 terms written down by LW.

Thus we have deduced all the relationships between $c_2$ and $c_3$ ascribed by LW to string duality, for all values of D, by purely classical
analysis. Not surprisingly, they also reproduce the truncated Arvis spectrum. Therefore the particular values of $c_2,c_3$ already saturate
the $\frac{1}{R^3}$ terms seen in the numerical simulations of Hari Dass and Pushan Majumdar. Does it mean that the effective string theory
has only the Nambu-Goto action? There is a priori no reason for that and even higher order terms with the same symmetry content as Nambu-Goto
action are certainly possible, and will in fact be explicitly constructed later in the chapter. This only means that there are no further
corrections to the spectrum at the $\frac{1}{R^3}$ level to what is given by the Nambu-Goto action. We will establish these results in
the Polchinski-Strominger effective string theory. This is done in the next section where it will also be shown that potentially there can be
order $\frac{1}{R^3}$ corrections to the spectrum even if there are no candidate actions at this level over and above the PS action.,

It is clear that in the static gauge, even just the Nambu-Goto action leads to an infinite number of higher order terms. This is a nuisance,
as in each order in powers of $\frac{1}{R}$, one has to disentangle the effects of the Nambu-Goto term to understand genuinely new additions
to the actions. As in most perturbation theory calculations, this quickly becomes very unwieldy. We will see in the next section that in
the covariant gauge characterising Polchinski-Strominger(PS) theory \cite{pseff1991}, the Nambu-Goto action only produces the leading 
quadratic term.
\section{Polchinski-Strominger(PS) Effective String Theory}
\label{sec:psest}
Introducing the composite operator 
\be
\label{eq:ch22-metriceq}
g_{\alpha\beta}\,=\,\partial_\alpha\,X^\mu\partial_\beta\,X_\mu
\ee
with $\xi^\alpha$ being the coordinates of the world sheet, it is easy to see that under general coordinate transformations of the 
the world-sheet coordinates $\xi^\alpha$, $g_{\alpha\beta}$ indeed transforms as a metric tensor provided $X^\mu$ transform as scalar fields.
We call it a \emph{metric substitute} and in this particular case, it has the meaning as the \emph{induced metric} on the world-sheet. Thus 
the Nambu-Goto action eqn.(\ref{eq:ch22-nambugoto}) is indeed invariant under the world-sheet general coordinate transformations.

Just as in the case of the static gauge, this invariance also permits a \emph{covariant gauge}, exactly as in the case of string theory (for
details please see chapter 17). Taking the world-sheet coordinates to be $(\tau,\sigma)$ and introducing the light-cone coordinates
$\tau^\pm\,=\,\tau\pm\sigma$, this covariant gauge, also called the conformal gauge, is fixed by the conditions
\be
\label{eq:ch22-conformgauge}
g_{++}(\tau^\pm)\,=\,g_{--}(\tau^\pm)\,=\,0
\ee
The only remaining component of the metric substitute is $g_{+-}=g_{-+}=\partial_+X\cdot\partial_-X$. The Nambu-Goto action then takes the
simple looking form (we have chosen to express it the same way as PS have; the string tension is $\frac{1}{4\pi\,a^2}$, with $a$ having 
dimensions of length):
\be
\label{eq:ch22-pszero}
S_0^{PS}\,=\,\frac{1}{4\pi\,a^2}\int\,d\tau^+d\tau^-\,\partial_+X^\mu\partial_-X_\mu
\ee
This is the amazing property of the conformal gauge: the Nambu-Goto action is equivalent to a single, quadratic action, unlike the static
gauge where it produced an infinite series of actions of increasing dimensions. In chapter 17, we had called this the Nielsen-Susskind action,
and also noted how Nambu had preferred the area action as being geometrical. But now we see that what he had rejected is just as geometrical
as what he had preferred, one being just a gauge-fixed version of the other. 
\subsection{Leading Order Analysis of PS Effective Actions}
\label{subsec:ch22-psleading}
Now we will analyse this leading order effective action, both to establish the general methodology as well as to rederive the Arvis spectrum
in the conformal gauge. While this follows Polchinski and Strominger \cite{pseff1991}, the spirit is somewhat different as we shall not
restrict to a perturbative expansion in $\frac{1}{R}$; instead we shall work out the exact spectrum, to all orders.

The equations of motion (Euler-Lagrange)(EOM) following from this action are:
\be
\label{eq:ch22-pszeroeom}
\partial_+\partial_-\,X^\mu\,=\,0
\ee
The classical ground state, analogous to the rigid string solution of LW is expressed as
\be
\label{eq:ch22-pszer0cla}
X_{cl}^\mu(\tau^\pm)\,=\,e_+^\mu\,R\,\tau^+\,+\,e_-^\mu\,R\,\tau^-
\ee
where $e_\pm^\mu$ are constant vectors. Actually, the most general solutions of the EOM are
\be
\label{eq:ch22-pszerogen}
X^\mu(\tau^\pm)\,=\,F^\mu(\tau^+)\,+\,G^\mu(\tau^-)
\ee
But $S_0^{PS}$, even after the gauge fixing, still has \emph{residual} local invariances under 
$\tau^+\rightarrow\,\tau^+\,+\,\epsilon^+(\tau^+)$ and $\tau^-\rightarrow\,\tau^-\,+\,\epsilon^-(\tau^-)$, with the corresponding
transformations of $X^\mu$:
\be
\label{eq:ch22-conftra}
\delta_+\,X^\mu\,=\,\epsilon^+(\tau^+)\,\partial_+\,X^\mu\quad\quad \delta_-\,X^\mu\,=\,\epsilon^-(\tau^-)\,\partial_-\,X^\mu 
\ee
This allows $F^\mu,G^\mu$ to be chosen as in eqn.(\ref{eq:ch22-pszerogen}).

Following PS, we also choose to analyse closed strings only though the numerical determinations were done for open strings. However, as
LW have discussed, there is a duality between them. Also, we need not consider issues concerning boundary terms in the action. Thus, all
solutions, and the classical ones in particular, satisfy the periodicity conditions
\be
\label{eq;ch22-psperiod}
X^\mu(\tau,\sigma+2\pi)\,=\,X^\mu(\tau,\sigma)\,+\,2\pi\,R\,\delta_1^\mu
\ee
where the string is taken to lie along the $X^1$-direction. This immediately leads to $e_+^\mu\,-\,e_-^\mu\,=\,\delta_1^\mu$.
The classical constraints of eqn.(\ref{eq:ch22-conformgauge}) yield $e_+\cdot\,e_+\,=\,e_-\cdot\,e_-\,=\,0$ i.e. these are null vectors. 
Combining this with the periodicity condition yields $e_+\cdot\,e_-\,=\,-\frac{1}{2}$.

The conserved energy-momentum tensors are
\be
\label{eq:ch22-psenmom}
T_{--}\,=\,-\frac{1}{2a^2}\,\partial_-X\cdot\partial_-X\quad\quad\, T_{++}\,=\,-\frac{1}{2a^2}\,\partial_-X\cdot\partial_-X
\ee
The fluctuations are now characterized by $Y^\mu$ given by $X^\mu\,=\,X_{cl}^\mu\,+\,Y^\mu$. In terms of these, $T_{--}$ takes the 
form (analogous expression for $T_{++}$ with $e_\rightarrow\,e_+,\partial_-\rightarrow\,\partial_+$):
\be
\label{eq:ch22-psflucenmom}
T_{--}\,=\,-\frac{R}{a^2}\,e_-\cdot\partial_-Y\,-\,\frac{1}{2a^2}\,\partial_-Y\cdot\partial_-Y
\ee
and likewise for $T_{++}$. Next, one introduces the mode expansions
\be
\label{eq:ch22-psmodes}
\partial_-Y\,=\,a\,\sum_{m=-\infty}^\infty\,\alpha_m^\mu\,e^{-i\,m\,\tau^-}\quad\quad\, \partial_+Y\,=\,a\,\sum_{m=-\infty}^\infty\,{\tilde{\alpha}}_m^\mu\,e^{-i\,m\,\tau^+}
\ee
The oscillator algebra is given by
\be
\label{eq:ch22-psoscalg}
[\alpha_m^\mu,\alpha_n^\nu]\,=\,m\,\eta^{\mu\nu}\,\delta_{m+n,0}\quad\quad\, [{\tilde{\alpha}}_m^\mu,{\tilde{\alpha}}_n^\nu]\,=\,m\,\eta^{\mu\nu}\,\delta_{m+n,0}
\ee
These are the same as what was encountered for free closed strings in chapter 17 and for the Shapiro-Virasoro(SV) model in chapter 16. The
Virasoro generators are again defined in terms of Fourier expansions of $T_{--}$ and $T_{++}$. The result is
\be
\label{eq:ch22-psvirgen}
L_n\,=\,\frac{R}{a}\,e_-\cdot\alpha_n\,+\,\frac{1}{2}\,\sum_{m=-\infty}^\infty\,:\alpha_{n-m}\cdot\alpha_m:
\ee
As in the earlier discussion of string theories, $:\ldots:$ denotes normal ordering. 
Likewise for ${\tilde{L}}_n$ with $e_-\rightarrow\,e_+$ and $\alpha_n\rightarrow\,{\tilde{\alpha}}_n$. It is easy to verify the Virasoro
algebras in both sectors to be
\be
\label{eq:ch22-psviralg}
[L_m,L_n]\,=\,(m-n)\,L_{m+n}\,+\,\frac{D}{12}(m^3-m)\,\delta_{m+n,0} 
\ee
likewise for ${\bar{L}}_n$. The additional $R$-dependent term in the Virasoro 
generators does not contribute to the central charge by virtue of 
$e_-\cdot\,e_-\,=\,e_+\cdot\,e_+\,=\,0$. All these manipulations are exactly as in string theories as discussed in chapter 17.
Therefore, the critical dimension remains 26.
 
As before, the quantum ground state is $|k,k;0\rangle$ which is an eigenstate of both $\alpha_0^\mu,{\tilde{\alpha}}_0^\mu$ with the same
eigenvalue $a\,k^\mu$. The physical state conditions are $L_0\,=\,{\tilde{L}}_0\,=\,1$ and $L_n\,=\,{\tilde{L}}_n\,=\,0$, for $n\ge\,2$. By
taking sum and difference of the $L_0,{\tilde{L}}_0$  conditions, and using the eigenvalue conditions, along with the periodicity condition 
$e_+^\mu\,-\,e_-^\mu\,=\,\delta_1^\mu$, it is easy to arrive at
\be
\label{eq:ch22-physgrnd}
k^1\,=\,0\quad\quad\,k^2\,+\,\frac{R}{a^2}(e_+\,+\,e_-)\cdot\,k\,=\,\frac{2}{a^2}
\ee
The total momentum of the string is (see section \S\,4.2 later)
\be
\label{eq:ch22-strmom}
p^\mu\,=\,\frac{R}{2a^2}\,(e_+^\mu\,+\,e_-^\mu)\,+\,\frac{1}{2a}(\alpha_0^\mu\,+\,{\tilde{\alpha}}_0^\mu)
\ee
which for the ground state is given by $p_{grnd}^\mu\,=\,\frac{R}{2a^2}(e_+\,+\,e_-)^\mu\,+\,k^\mu$. Therefore, the total rest energy of the ground state is
\be
\label{eq:ch22-grndmass}
(-p^2)^{\frac{1}{2}}\,=\,\sqrt{(\frac{R}{2a^2})^2\,-\,k^2\,-\frac{R}{a^2}(e_+\,+\,e_-)\cdot\,k}
\ee
which, on using the second of eqn.(\ref{eq:ch22-physgrnd}) becomes
\be
\label{eq:ch22-grndmass2}
(-p^2)^{\frac{1}{2}}|_{grnd}\,=\,\frac{R}{2a^2}\,\sqrt{1\,-\,2\,(\frac{2a}{R})^2}
\ee
which is the Arvis result for closed strings, but for $D=26$! The point is that even though eqn.(\ref{eq:ch22-arvisfull}) is made to look
like a function of D, it is only meaningful in $D=26$, as demonstrated by Arvis himself on requiring consistency with rotational invariance
(see chapter 17 for the details). Therefore, it is only legitimate to use the Arvis potential for $D=26$.

The big question then is the legitimacy of using the expression for the L\"uscher term and the truncated Arvis potential for $D\ne\,26$. PS
give a resolution of this and we discuss it in the next section.
\section{PS Effective String Theories for all D}
\label{sec:ch22-PSeff}
Polchinski and Strominger \cite{pseff1991} constructed effective string theories that are valid in all dimensions D, not just $D=26$. 
Let us recall the origin of the $D=26$ difficulty in string theories(see chapter 17 for details). One way of understanding this is through 
the \emph{Conformal Anomaly} (for a brief discussion of anomalies the reader is referred to chapter 18). Only in 26 dimensions is this 
anomaly absent. However, Polyakov \cite{polyasubcr1981} had shown that strings can be formulated in $D\,<\,26$ dimensions. This is called
Polyakov subcritical string theory. 

As an alternative to the Nambu-Goto action, Polyakov introduces an action, by enlarging the field content to also include an 
\emph{intrinsic metric} $h_{\alpha\beta}$ on the world sheet transforming exactly as the induced metric of eqn.(\ref{eq:ch22-metriceq}) under world-sheet
general coordinate transformations:
\be
\label{eq:ch22-polyaact}
S_{polya}\,=\,\int\,d^2\xi\,\sqrt{h}\,h^{\alpha\beta}\,\partial_\alpha\,X\cdot\partial_\beta\,X
\ee
where $h$ is the determinant of $h^{\alpha\beta}$. With the inclusion of the intrinsic metric, invariance under world-sheet general coordinate
invariance now becomes exactly akin to that in General Relativity theory, unlike that in the case of the Nambu-Goto action, as 
already remarked. However, the Polyakov action is also invariant under \emph{local Weyl-scalings}:
\be
\label{eq:ch22-weylscaling}
h_{\alpha\beta}\rightarrow\,\omega(\xi)\,h_{\alpha\beta}\quad\quad\,X^\mu\rightarrow\,X^\mu
\ee
This forms the symmetry content of the string theory now. Polyakov analysed this theory in the path-integral approach. We will have lot more
to say about this class of actions shortly.

Fujikawa had 
shown \cite{fujimeas1979}, in the context of chiral anomalies, that in the path-integral formulation, the entire anomaly resides in the
path-integral measure. Now in $D\ne\,26$, the Weyl-anomaly also resides in the path-integral measures for the functional integrations
over $X^\mu$ and $h_{\alpha\beta}$. By utilizing general coordinate transformations, every two dimensional metric can be brought into a
\emph{conformally flat} metric and the only degree of freedom left is the conformal factor $e^\phi(\xi)$. Now the measure for 
$\phi$-integrations being no longer invariant, these integrations induce an action for the additional degree of freedom $\phi$. We refer
the reader to \cite{polyasubcr1981} for this very deep aspect and simply quote the result for this so called \emph{Liouville Action}:
\be
\label{eq:ch22-liouville}
S_L\,=\,\frac{26-D}{48\pi}\int\,d^2\xi\,\partial_+\phi\,\partial_-\phi
\ee
The way to understand this beautiful work of Polyakov is to realise that the additional degree of freedom makes the right additional
contribution to the central charge of the Virasoro algebra so as to completely cancel the conformal anomaly in every D. Viewed as a fundamental
theory, the Liouville theory also has some problems like the so called $c=1$ \emph{barrier}, but as an effective theory, hopefully, this
will not be a problem. Actually, the Liouville action also has a difficult to handle, $e^{\mu\,\phi}$ part, which PS have ignored. For the sake of formulating an
effective theory, as PS do, this omission is permissible. Not surprisingly, this action is proportional to $(26-D)$. Without the
exponential term, the action is invariant under shifts of the $\phi$-field. Now Polchinski and Strominger introduce an effective action
only for $X^\mu$ by equating $h_{+-}$ of the conformal gauge to the induced metric $g_{+-}$ introduced earlier i.e by choosing $e^\phi\,=\,g_{+-}$. Calling this L, the effective action one obtains is
\be
\label{eq:ch22-polyaliou}
S_{PL}\,=\,\frac{26-D}{48\pi}\int\,d\tau^+\,d\tau^-\,\frac{\partial_+\,L\partial_-\,L}{L^2}
\ee
exactly invariant under the same conformal transformations of eqn.(\ref{eq:ch22-conftra}) that left $S_0^{PS}$ invariant. This action is clearly
non-polynomial in general. As configurations can easily be found for which the denominator can vanish i.e. $\partial_+X\cdot\partial_-X\,=\,0$,
it is also singular in general. But as an effective action for \emph{long} strings whose classical configuration satisfies 
$L\,=\,-\frac{R^2}{2}$, it should be acceptable (unless proven otherwise).

Instead of analysing 
their effective action in this beautiful form, PS now proceed to make things superficially confusing by throwing away all terms proportional
to the leading Euler-Lagrange equations of motion(EOM) from $S_0^{PS}$ i.e. $\partial_+\partial_-\,X^\mu\,=\,0$. They do this by suitable
\emph{field redefinitions} of $X^\mu$ resulting in the somewhat simpler action
\be
\label{eq:ch22-psact}
S_{PS}\,=\,\frac{\beta}{4\pi}\,\int\,d\tau^+\,d\tau^-\,\frac{\partial_+^2\,X\cdot\partial_-X\,\partial_-^2\,X\cdot\partial_+X}{(\partial_+X\cdot\partial_-X)^2}
\ee
In addition to throwing away all terms proportional to the leading order EOM, they also throw caway all terms proportional to the leading
order constraints $\partial_+X\cdot\partial_+X\,=\,0\,=\,\partial_-X\cdot\partial_-X$ (see \cite{pseff1991} for details). They also kept the 
coefficient of the action as a free parameter $\frac{\beta}{4\pi}$ to be determined later. 

But the price to be paid for this field redefinition is that the transformation laws under which the simplified action(plus $S_0^{PS}$) are
invariant looks like a real mess! PS had to determine this iteratively i.e. by truncating the action to a certain order in $\frac{1}{R}$,
the transformation laws were also determined to a certain suitable order, a decidedly tedious procedure. 
Neglecting terms of order $\frac{1}{R^3}$, equivalently keeping only terms upto order $\frac{1}{R^2}$($O(R^{-3})$), they determined the
transformation laws 
\be
\label{eq:ch22-pstra}
\delta\,X^\mu\,=\,\epsilon^-(\tau^-)\,\partial_-X^\mu\,-\,\frac{\beta\,a^2}{2}\,\partial_-^2\epsilon^-(\tau^-)\,\frac{\partial_+\,X^\mu}{\partial_+X\cdot\partial_-X}
\ee
It is to be appreciated that the orders of truncations in the action and the transformation laws should be self-consistent. It is an elementary
exercise to work out that if terms of order $R^{-N}$ are kept in the action, terms to the same order must be retained in the transformation
laws also. In their original analysis of \cite{pseff1991}, they retained terms of order $R^{-2}$ in both of them(i.e. $O(R^{-3})$). Obviously,
the closure of the algebra of transformations will also be valid only upto some order.

When we started working on the PS theory\footnote{This author became aware of the PS approach from remarks by Professor Julius Kuti as 
conveyed by Professor Apoorva Patel.}, we took on face value their statement, immediately after their eqn.(5) where they had stated the
Liouville action of eqn.(\ref{eq:ch22-liouville}), that substitution of the induced metric in place of $e^\phi$ gives their action of
eqn.(\ref{eq:ch22-psact}). It is their transformation law that baffled this author. The clue finally came on his working out the algebra
of their transformation laws of eqn.(\ref{eq:ch22-pstra}) to be
\be
\label{eq:ch22-pstralgebra}
[\delta_-^{PS}(\epsilon_1^-),\delta_-^{PS}(\epsilon_2^-)]\,=\,\delta_-^{PS}(\epsilon_{12}^-)\,+\,O(R^{-4})
\ee
with $\epsilon_{12}^-\,=\,\epsilon_1^-\partial_-\epsilon_2^-\,-\,\epsilon_2^-\partial_-\epsilon_1^-$. The algebra of the conformal
transformations of eqn.(\ref{eq:ch22-conftra}) is
\be
\label{eq:ch22-conftralgebra}
[\delta_-^{0}(\epsilon_1^-),\delta_-^{0}(\epsilon_2^-)]\,=\,\delta_-^{0}(\epsilon_{12}^-)
\ee
which is exact i.e. valid to \emph{all} orders in $R$. This meant that the PS transformation laws were just conformal transformations in disguise! This also meant that the strange-looking PS 
transformations and conformal transformations ought to be related by field redefinitions, to some order of accuracy. The author then by
hand constructed such a field redefinition and was surprised that the Liouville action could be recovered. Furthermore, it could be
seen that the Liouville action itself was \emph{exactly} invariant under eqn.(\ref{eq:ch22-conftra}), an important point that had not been
stressed sufficiently by PS.

Now we present the original analysis of PS of their effective action which removes the conformal anomaly in every D. It centers around the
modifications to the Virasoro generators and their algebra due to the added effective action. For this, one needs to construct the conserved
energy-momentum tensors $T_{--},T_{++}$ through Noether's theorem though care has to be exercised as conformal invariance is local. 
The new \emph{On-shell} energy-momentum tensors are (i.e. after throwing away terms proportional to the EOM's):
\br
\label{eq:ch22-psenmom2}
T_{--}^{PS}\,&=&\,-\frac{1}{2a^2}\partial_-X\cdot\partial_-X
+\frac{\beta}{2L^2}(L\partial_-^2\,L-(\partial_-L)^2\nonumber\\
\,&+&\,\partial_-X\cdot\partial_-X\,\partial_+^2X\cdot\partial_-^2X
\,-\,\partial_+L\,\partial_-X\cdot\partial_-^2X)
\er
It is a bit tedious to compare this with what Drummond has given in his eqn.(2.12) \cite{drumm2004}. We will soon compare our corresponding expressions for
fluctuations to order $R^{-2}$(the relevant order to compute the spectrum to order $R^{-3}$).
As 
the leading term in the transformation law grows as $R$, the leading term of the energy momentum tensor also grows like $R$, and as the 
action is 
$O(R^{-3})$, one can only get $T_{\pm\pm}$ to $O(R^{-2})$. In terms of the fluctuations $Y^\mu$ introduced earlier, the energy momentum tensors
now read (this should be compared with the result of the leading order analysis in eqn.(\ref{eq:ch22-psflucenmom})):
\be
\label{eq:ch22-psflucenmom2}
T_{--}\,=\,-\frac{R}{a^2}\,e_-\cdot\partial_-Y\,-\,\frac{1}{2a^2}\,\partial_-Y\cdot\partial_-Y\,-\,\frac{\beta}{R}\,e_+\cdot\partial_-^3Y+\ldots
\ee
It is clear, just by inspection, that the interplay between the first and last terms has the effect of shifting the central charge by $12\beta$
i.e. from $D$ to $D\,+\,12\beta$! PS show this by working out the \emph{Operator Product Expansion}(OPE) for $T_{--},T_{++}$. 

It can equally well be shown by constructing the Virasoro generators $L_n$ by Fourier-analysing the energy momentum tensors. Following the
standard methods discussed earlier, the generators are found to be
\be
\label{eq:ch22-psvirgen2}
L_n\,=\,\frac{R}{a}\,e_-\cdot\alpha_n\,+\,\frac{1}{2}\,\sum_{m=-\infty}^\infty\,:\alpha_{n-m}\cdot\alpha_m:\,+\,\frac{\beta}{2}\,\delta_{n,0}\,-\,\frac{a\beta\,n^2}{R}\,e_+\cdot\alpha_n\,+\,O(R^{-2})
\ee
and likewise for ${\tilde{L}}_n$ in terms of ${\tilde{\alpha}}_n$. It is again a straightforward exercise to show that they satisfy the
algebra
\be
\label{eq:ch22-psviralg2}
[L_m,L_n]\,=\,(m-n)\,L_{m+n}\,+\,\frac{D+12\beta}{12}(m^3-m)\,\delta_{m+n,0} 
\ee
likewise for ${\bar{L}}_n$. The additional contributions to the central extension term come from two important sources: a) the cross terms from the first and last terms
in $L_n$ where the factors of R exactly compensate each other, and, b) the shift of $L_0$ by $\frac{\beta}{2}$. As already discussed in Chapter
16, the general structure of this term follows from very general considerations like Jacobi identities, only the coefficient needs some
explicit calculations.

The implications of this modification for the criticality of strings is dramatic. Instead of the earlier requirement of $D=26$, the new
requirement is $D+12\beta\,=\,26$ which can be satisfied for every D as long as $\beta$ is chosen to be the critical value 
$\beta_c\,=\,-\frac{D-26}{12}$. Remarkably, the coefficient $\frac{\beta}{4\pi}$ of the action takes the same value as in the original
Polyaklov-Liouville theory. This leads to the final expression for the Virasoro generators:
\be
\label{eq:ch22-psvirgen3}
L_n\,=\,\frac{R}{a}\,e_-\cdot\alpha_n\,+\,\frac{1}{2}\,\sum_{m=-\infty}^\infty\,:\alpha_{n-m}\cdot\alpha_m:\,+\,\frac{\beta_c}{2}\,\delta_{n,0}\,-\,\frac{a\beta_c\,n^2}{R}\,e_+\cdot\alpha_n\,+\,O(R^{-2})
\ee
In what follows we shall always mean $\beta_c$ even if we use $\beta$.

We now analyse the ground state energy; the case of excited states is straightforward to generalize. The quantum ground state $|k,k;0\rangle$
is a simultaneous eigenstate of both $\alpha_0^\mu$ and ${\tilde{\alpha}}_0^\mu$ with the eigenvalue $a\,k^\mu$. The sum and difference of
the conditions $L_0\,=\,{\tilde{L}}_0\,=\,1$ now yields
\be
\label{eq:ch22-physgrnd2}
k^1\,=\,0\quad\quad\,k^2\,+\,\frac{R}{a^2}(e_+\,+\,e_-)\cdot\,k\,=\,\frac{2\,-\,\beta_c}{a^2}
\ee
To this order, the ground state momentum still remains $p_{grnd}^\mu\,=\,\frac{R}{2a^2}(e_+^\mu\,+\,e_-^\mu)\,+\,k^\mu$(this will be 
explained in detail shortly). The resulting ground state mass becomes
\be
\label{eq:ch22-grndmass3}
(-p^2)^{\frac{1}{2}}|_{grnd}\,=\,\frac{R}{2a^2}\,\sqrt{1\,-\,\frac{D-2}{12}\,(\frac{2a}{R})^2}
\ee
But this analysis is only valid to $O(R^{-2})$ and therefore self-consistency requires that this be truncated  to this order, yielding
\be
\label{eq:ch22-psluscher}
V(R)\,=\,\frac{R}{2a^2}\,-\,\frac{D-2}{12}\,\frac{1}{R}\,+\ldots
\ee
giving the universal L\"uscher term for closed strings, but now for all D. This vindicates the use of the L\"uscher term in interpreting 
numerical simulation results for $D\ne\,26$. Now we turn to showing how the truncated Arvis potential, valid for all D, also emerges from
the PS theory at the next non-trivial higher order.
\subsection{Order $R^{-3}$ corrections to the spectrum.}
\label{subsec:ch22-pstruncarvis}
Now we address the important issue of possible terms in the spectrum at order $R^{-3}$ i.e. $O(R^{-4})$. The truncated Arvis potential to 
which we had claimed numerical evidence in $D=3$ SU(2) and $D=4$ SU(3) \cite{ndhpushan2005,ndhpushan2006,ndhpushan2007,ndhpushan2008} is
of this type. The truncated Arvis spectrum is the analogous result for excited states. To apply the PS formalism to this end requires
going beyond their analysis. Following their methodology would require on the one hand identification of possible additional terms to the
action, and on the other hand find the modifications to their transformation laws that would keep the total action invariant. PS themselves
stated, without any elaboration, that the next such terms in the action are of order $R^{-4}$. Since everything depends on the correctness
of this assertion, a more systematic analysis is called for. There is also the related issue of extending the transformation laws.

This problem was independently solved by Drummond in \cite{drumm2004}, and by this author and Peter Matlock in their work \cite{ndhmat2006}.
Both the approaches were technically almost identical. While Hari Dass and Matlock had shown that there were no additional terms in the action 
at $R^{-3}$ order over and above what one obtains on expanding the PS-action to that level, they had also claimed that there would be such
additions at $R^{-4}$ and $R^{-5}$ orders, Drummond had claimed that in fact the possible corrections would only be at order $R^{-6}$.
Hari Dass and Matlock had criticised the claims made by Drummond regarding terms of order higher than $R^{-3}$, and in response 
Drummond \cite{drummreply}
had given a more explicit and systematic treatment which upheld his earlier claims. So indeed the potential corrections would, rather
remarkably, only appear at order $R^{-6}$ and higher. Shortly afterwards, Hari Dass and Matlock gave a systematic method for constructing
all action candidates that had the virtue of keeping the transformation laws fixed as those that left the leading action $S_0^{PS}$ invariant. In
other words, their construction, which they called a \emph{Covariant Calculus for Effective String Theories} \cite{covcal2007,covcal2014}
enables one to construct actions that are invariant to all orders in $R^{-1}$ under the same fixed transformation laws as given by
eqn.(\ref{eq:ch22-conftra}), without the need to keep adjusting the transformation laws every time a new term was added to the action as
per the methods of PS. We will discuss the covariant calculus in detail shortly. The upshot of that calculus was a straightforward
vindication of Drummond's claims about the only corrections being at order $R^{-6}$ and higher, but of the four candidate actions proposed
by Drummond, only two specific combinations appeared. 

PS themselves had given a recipe for constructing the additional terms to their effective action. In essence, it involved constructing
actions that transform like (1,1) in a naive sense meaning the net number of + and - indices are one each. This naive criterion should be
distinguished from quantities actually transforming as (1,1) under transformations analogous to the PS-transformation laws, but the
complication now is that the forms of these transformation laws actually depend on the precise choice of terms in the action. The naive
criterion is necessary but not sufficient. In fact, the PS lagrangian is a case in point, it is only (1,1) in the naive sense and does not
strictly transform as a (1,1) tensor; nevertheless, the PS-action is invariant. This distinction will be made clearer when we discuss the
covariant calculus. 

The typical form of such terms is one with a numerator with certain number of (+,-) indices and a denominator with the correct numbers of
these indices to provide a net (1,1) term. The denominators should not become singular even in the limited context of fluctuations
around a classical background of a rigid string. That only leaves powers of $L\,=\,\partial_+X\cdot\,\partial_-X$. This is certainly so
for the PS-lagrangean.

In the next part of the recipe, PS advocate throwing away all terms proportional to the leading order EOM's as they can be eliminated
through field redfinitions albeit at the cost of changing the transformation laws. Finally, PS also advocate throwing away terms proportional
to the leading order constraints $\partial_{\pm}X\cdot\partial_{\pm}X$. Needless to say, this recipe gets rapidly unwieldy with each increasing
order.

At first, both Drummond, and, Hari Dass and Matlock(HM) followed this procedure. Initially, the latter had claimed potential corrections at order $R^{-4},R^{-5}$, but Drummond in \cite{drummreply} showed that all these are actually equivalent to order $R^{-6}$ terms after partial integrations.
We refer the reader to these sources for details but the important punchline agreed to by both the groups is \emph{the absence of terms at
order $R^{-3}$} over and above what one already obtained from expanding the PS-action to this order. HM also showed that the PS-transformation
laws are actually good to terms including $R^{-3}$ terms.

The consequence of this is that the on-shell energy-momentum tensors can be computed to include terms of order $R^{-2}$ by simply expanding
eqn.(\ref{eq:ch22-psenmom2}). The result (with a similar expression with $+\leftrightarrow\,-$) is
\br
\label{eq:ch22-psflucenmom3}
& &T_{--}\,=\,-\frac{R}{a^2}\,e_-\cdot\partial_-Y\,-\,\frac{1}{2a^2}\partial_-Y\cdot\partial_-Y\,-\,\frac{\beta}{R}\,e_+\cdot\partial_-^3Y
-\frac{\beta}{R^2}\,\{2(e_+\cdot\partial_-^2Y)^2\nonumber\\
&+&2e_+\cdot\partial_-^3Y(e_+\cdot\partial_-Y+e_-\cdot\partial_+Y)+2e_-\cdot\partial_-^2Ye_-\cdot\partial_+^2Y+\partial_+Y\cdot\partial_-^3Y\}
\er
This is in complete agreement with that of Drummond as in his eqn.(2.16). However, he uses somewhat different techniques in his next steps. We
shall continue to follow \cite{ndhmat2006}.

This energy-momentum is conserved on-shell i.e.
\be
\label{eq:ch22-enmomcons}
\partial_+\,T_{--}\,=\,0
\ee
Uptil now $T_{--}$ only involved derivatives of $Y^\mu$ wrt $\tau^-$ and this conservation was obvious in view of the EOM 
$\partial_+\partial_-\,Y\,=\,0$. But now, at order $R^{-2}$, there are two terms that involve derivatives of Y wrt $\tau^+$ and it is not 
immediately obvious how the conservation law follows. For this we need to examine the EOM to order $R^{-3}$. We quote the result for this
EOM and refer the interested reader to \cite{ndhmat2006}:
\br
\label{eq:ch22-eom3rd}
& &\frac{2}{a^2}\partial_{+-}Y^\mu\,=\,-4\frac{\beta}{R^2}e_+\cdot\partial_+^2\partial_-^2\,Y\,e_-^\mu -\frac{4\beta}{R^3}\{\partial_+^2[\partial_-^2Y^\mu(e_+\cdot\partial_-Y+e_-\cdot\partial_+Y)]\nonumber\\
&+&\, \partial_-^2[\partial_+^2Y^\mu(e_+\cdot\partial_-Y+e_-\cdot\partial_+Y)]\,+\,4\,e_+^\mu\partial_-(\partial_+^2Y\cdot\partial_-^2Y)\,+\,4\,e_-^\mu\partial_+(\partial_+^2Y\cdot\partial_-^2Y)\}\nonumber\\
\er
This can be simplified considerably on noting that all terms involving $\partial_+\partial_-Y$ and it's derivatives will actually be higher
order in $R^{-1}$ than relevant. The simplified form of EOM becomes
\br
\label{eq:ch22-eom3rd2}
\frac{2}{a^2}\,\partial_{+-}Y^\mu\,&=&\,\frac{4\beta}{R^3}\{e_-^\mu\partial_+(\partial_+^2Y\cdot\partial_-^2Y)\,+\,e_+^\mu\partial_-(\partial_+^2Y\cdot\partial_-^2Y)\nonumber\\
&-&\,\partial_-^2(\partial_+^2Y^\mu\,e_+\cdot\partial_-Y)\,-\,\partial_+^2(\partial_-^2Y^\mu\,e_-\cdot\partial_+Y)\}
\er 
This can be solved iteratively by writing $Y^\mu\,=\,Y_0^\mu\,+\,Y_1^\mu$ with $Y_0^\mu$ being the solution to the leading order EOM,leading
to
\br
\label{eq:ch22-eomred}
\frac{2}{a^2}\,\partial_{+-}Y_1^\mu\,&=&\,\frac{4\beta}{R^3}\{e_-^\mu\partial_+(\partial_+^2Y_0\cdot\partial_-^2Y_0)\,+\,e_+^\mu\partial_-(\partial_+^2Y_0\cdot\partial_-^2Y_0)\nonumber\\
&-&\,\partial_-^2(\partial_+^2Y_0^\mu\,e_+\cdot\partial_-Y_0)\,-\,\partial_+^2(\partial_-^2Y_0^\mu\,e_-\cdot\partial_+Y_0)\}
\er
This can be readily integrated to give
\be
\label{eq:ch22-eomint}
\frac{2}{a^2}\partial_-Y^\mu\,=\,\frac{4\beta}{R^3}\{e_+^\mu\partial_+Y_0\cdot\partial_-^2Y_0\,+\,e_-^\mu\partial_-^2Y_0\cdot\partial_+^2Y_0\,-\,\partial_-^2Y_0^\mu\,e_-\cdot\partial_+^2Y_0\,-\,\partial_=Y_0^\mu\,e_+\cdot\partial_-^3Y_0\}
\ee
On substituting this into the first term $-\frac{R}{a^2}e_-\cdot\partial_-Y$ of $T_{--}$, one sees the remarkable cancellation of all terms
with $+$-derivatives, leaving behind an expression for $T_{--}$ that is only built with $-$derivatives of $Y_0$, satisfying the conservation 
law in a straightforward manner. This is the purely holomorphic
representation of $T_{--}$(in \cite{ndhmat2006} there was a typo and we had used \emph{holonomic}, a meaningless phrase in this context!). Now 
it is also clear why we had to solve EOM to order $R^{-3}$ to get this cancellation to work at order $R^{-2}$. The final expression for 
$T_{--}$ is:
\br
\label{eq:ch22-psR3enmomfin}
T_{--}\,&=&\,-\frac{R}{a^2}\,e_-\cdot\partial_-Y_0\,-\,\frac{1}{2a^2}\partial_-Y_0\cdot\partial_-Y_0\,-\,\frac{\beta}{R}e_+\cdot\partial_-^3Y_0\nonumber\\
&-&\,\frac{2\beta}{R^2}e_+\cdot\partial_-^3Y_0\,e_+\cdot\partial_-Y_0\,-\,\frac{2\beta}{R^2}(e_+\cdot\partial_-^2Y_0)^2
\er
This fully agrees with the Drummond's expression. With the mode expansion for $Y_0^\mu$ already discussed, the Virasoro generators at this
order are:
\br
\label{eq:ch22-virgenR3}
L_n\,&=&\,\frac{R}{a}e_-\cdot\alpha_n\,+\,\frac{1}{2}\sum_{m=-\infty}^\infty:\alpha_{n-m}\alpha_m:\,+\,\frac{\beta_c}{2}\delta_{n,0}\,
-\,\frac{a\beta_cn^2}{R}e_+\cdot\alpha_n\nonumber\\
&-&\,\frac{\beta_ca^2n^2}{R^2}\sum_{m=-\infty}^\infty\,:e_+\cdot\alpha_{n-m}e_+\cdot\alpha_m:\,+O(R^{-3})
\er
Agreeing with Drummond's formulae for the same.

Now, the additional modifications to $L_0,{\tilde{L}}_0$ actually vanish! Thus the spectrum of PS-effective string theories to order $R^{-3}$
is the same as that of the truncated Arvis potential, but now with the dramatic reinterpretation that it is valid for all D, and not just
$D=26$!
\subsection{Ground State Momentum Revisited}
\label{subsec:ch22-stringmom}
There is, however, a serious caveat to the above derivation. It depends on the tacit assumption that the ground state momentum does not
receive any corrections even at $R^{-3}$ order and the earlier expression $p_{grnd}^\mu\,=\,\frac{R}{2a^2}(e_+^\mu\,+\,e_-^\mu)\,+\,k^\mu$  
still holds. Drummond just states this without elaboration. As this is a crucial ingredient and also as the derivation of this has a lot
of pedagogical value, we outline the proof that the author had given in \cite{ndhmom2010}. In that work, both the PS-action as well as the
Polyakov-Liouville action of eqn.(\ref{eq:ch22-polyaliou}) have been analysed. We start by illustrating the case of $S_0^{PS}$.

The action $S_0^{PS}$ is clearly invariant under $\delta_b\,X^\mu(\sigma,\tau)\,=\,b^\mu$ where $b^\mu$ does not depend on the world-sheet
coordinates $(\sigma,\tau)$. Therefore this is a case of a global invariance and N\"other's theorem can be applied straightforwardly.
The trick to compute the corresponding conserved quantities is to consider a $(\sigma,\tau)$-dependent $b^\mu(\sigma,\tau)$. The 
variation of the action, which is no longer zero, will then be of the form
\be
\label{eq:ch22-bvar0}
\delta_b\,S_0^{PS}\,=\,\frac{1}{4\pi\,a^2}\,\int\,d\tau^+\,d\tau^-\,\{\partial_+b\cdot\partial_-X\,+\,\partial_+X\cdot\partial_-b\}
\ee
The conserved momentum densities are given by 
\be
\label{eq:ch22-noethermom}
\delta_b\,S_0^{PS}\,\equiv\,\int\,(p_+\cdot\partial_-b\,+\,p_-\cdot\partial_+b)
\ee
Therefore
\be
\label{eq:ch22-freemomden}
p_+^\mu\,=\,\frac{1}{4\pi\,a^2}\,\partial_+X^\mu\quad\quad\, p_-^\mu\,=\,\frac{1}{4\pi\,a^2}\,\partial_-X^\mu
\ee
The total conserved momentum is obtained by integrating $p_\tau$ over all $\sigma$(see chapter 17 for similar details in string theory). 
This is given by
\be
\label{eq:ch22-freemomdentau}
p_\tau^\mu(\sigma,\tau)\,=\,p_+^\mu\,+\,p_-^\mu
\ee
On expanding $X^\mu$ around the classical background $X_{cl}^\mu\,=\,R(e_+^\mu\tau^+\,+\,e_-^\mu\tau^-)$ i.e. $X^\mu\,=\,X_{cl}^\mu\,+\,Y^\mu$,
and using the mode expansion for $Y^\mu$ owing to the EOM $\partial_{+-}Y^\mu\,=\,0$
\be
\label{eq:ch22-leadmode}
Y^\mu(\sigma,\tau)\,=\,q^\mu\, +\,(a\alpha_0^\mu\tau^-\,+\,ia\sum_{m\ne\,0}\,\frac{\alpha_m^\mu}{m}\,e^{-im\tau^-}) +\,(a{\tilde{\alpha}}_0^\mu\tau^+\,+\,ia\sum_{m\ne\,0}\,\frac{{\tilde{\alpha}}_m^\mu}{m}\,e^{-im\tau^+})
\ee
It is easy to show that
\be
\label{eq:ch22-freemomdentau2}
p_\tau^\mu(\sigma,\tau)\,=\,\frac{R}{4\pi\,a^2}(e_+^\mu\,+\,e_-^\mu)\,+\,\frac{1}{4\pi\,a}(\alpha_0^\mu\,+\,{\tilde{\alpha}}_0^\mu)\,+\ldots
\ee
The $\ldots$ above integrate to zero, giving the expression for the total momentum (for closed strings)
\be
\label{eq:ch22-freemom}
p^\mu\,=\,\frac{R}{2a^2}(e_+^\mu\,+\,e_-^\mu)\,+\,\frac{1}{2a}(\alpha_0^\mu\,+\,{\tilde{\alpha}}_0^\mu)
\ee
After this warm-up exercise with the much simpler $S_0^{PS}$, we now turn to the case of the PS-action. Following the same procedures as in the case
of $S_0^{PS}$, the results for the PS-action are:
\br
\label{eq:ch22-psmomden}
\Delta\,{p^{ps}_+}^\mu\,&=&\,\frac{\beta}{4\pi}\{\frac{\partial_{++}X^\mu\,\partial_{--}X\cdot\partial_+X}{L^2}\,-\,\frac{2\,\partial_+X^\mu\,
\partial_{++}X\cdot\partial_-X\partial_{--}X\cdot\partial_+X}{L^3}\nonumber\\
&-&\,\partial_-(\frac{\partial_+X^\mu\partial_{++}X\cdot\partial_-X}{L^2})\}\nonumber\\
\er
with a similar expression for $\Delta\,{p^{ps}_-}^\mu$. As we shall see, a separation of $Y^\mu$ into $Y_0^\mu\,+\,Y_1^\mu$ as before is not even necessary, nor any mode expansions. Instead, the
general form of all functions of $(\tau^+,\tau^-)$ is
\be
\label{eq:ch22-generalY}
Y^\mu(\tau^+,\tau^-)\,=\,F^\mu(\tau^+)\,+\,G^\mu(\tau^-)\,+\,H^\mu(\tau^+,\tau^-)
\ee
where $H^\mu$ is such that it has no purely \emph{holomorphic}(i.e. functions of $\tau^-$ only) or \emph{anti-holomorphic}(functions of 
$\tau^+$ only) parts. From the EOM to order $R^{-3}$ it follows that (this can also be obtained from eqn.(\ref{eq:ch22-eomint}) on substituting
$Y_0\,=\,F\,+\,G$ and $Y_1\,=\,H$ and integrating it once more wrt to $\tau^-$):
\be
\label{eq:ch22-genH}
H^\mu\,=\,\frac{2\beta\,a^2}{R^3}\,[-e_-\cdot\partial_{++}F\,\partial_-G^\mu\,-\,e_+\cdot\partial_{--}G\,\partial_+F^\mu\,+\,e_+^\mu\,\partial_+F\cdot\partial_{--}G\,+\,e_-^\mu\,\partial_{++}F\cdot\partial_-G]  
\ee
Using these, the corrections to the momentum densities arising from the PS-action can be shown to be(with the short-hand notation $F_{++}\,=\,\partial_{++}F$ etc.)
\br
\label{eq:ch22-psmomden2}
\Delta\,{p^{ps}_+}^\mu\,&=&\,\frac{\beta}{2\pi\,R^3}\{e_-^\mu\,F_{+++}\cdot\,G_-\,-\,e_+^\mu\,F_{++}\cdot\,G_{--}\,-\,G_-^\mu\,e_-\cdot\,F_{+++}\,+\,F_{++}^\mu\,e_+\cdot\,G_{--}\}\nonumber\\
\Delta\,{p^{ps}_-}^\mu\,&=&\,\frac{\beta}{2\pi\,R^3}\{e_+^\mu\,G_{--+}\cdot\,F_+\,-\,e_-^\mu\,G_{--}\cdot\,F_{++}\,-\,F_+^\mu\,e_+\cdot\,G_{---}\,+\,G_{--}^\mu\,e_-\cdot\,F_{++}\}\nonumber\\
\er
Rather remarkably, these equations can be re-expressed as
\be
\label{eq:ch22-psmomdenimp}
\Delta\,{p^{ps}_+}^\mu\,=\,\frac{\beta}{2\pi\,R^3}\partial_+\,C^\mu\quad\quad\, \Delta\,{p^{ps}_-}^\mu\,=\,-\frac{\beta}{2\pi\,R^3}\partial_-\,C^\mu
\ee
with
\be
\label{eq:ch22-psmomdenimp2}
C^\mu\,=\,e_-^\mu\,F_{++}\cdot\,G_-\,-\,e_+^\mu\,F_+\cdot\,G_{--}\,+\,e_+\cdot\,G_{--}\,F_+^\mu\,-\,e_+\cdot\,F_{++}\,G_{--}^\mu
\ee
In other words, the corrections of order $R^{-3}$ to the momentum densities of the PS-action are of the so called \emph{improvement} type,
whose contribution to the total momentum vanishes identically! This is so because $\Delta{p^{ps}_\tau}^\mu$ has the form of a derivative
wrt $\sigma$ and $Y^\mu$ vanishes at the boundaries of the $\sigma$-integration. Thus the total momentum does not receive any corrections
of order $R^{-3}$ i.e. they are $O(R^{-4})$.
\section{Covariant Calculus For Effective String Theories}
\label{sec:ch22-covcaleff}
As already mentioned, the Polyakov-Liouville action
\be
\label{eq:ch22-polyaliou2}
S_{PL}\,=\,\frac{26-D}{48\pi}\int\,d\tau^+\,d\tau^-\,\frac{\partial_+\,L\partial_-\,L}{L^2}
\ee
is \emph{exactly} invariant under(induced by the infinitesimal transformations $\tau^\pm\rightarrow\,\tau^\pm\,-\,\epsilon^\pm(\tau^\pm)$)
\be
\label{eq:ch22-conftra2}
\delta_+\,X^\mu\,=\,\epsilon^+(\tau^+)\,\partial_+\,X^\mu\quad\quad \delta_-\,X^\mu\,=\,\epsilon^-(\tau^-)\,\partial_-\,X^\mu 
\ee
This should be contrasted with only the approximate invariance of eqn.(\ref{eq:ch22-psact}) under the approximate transformations of
eqn.(\ref{eq:ch22-pstra}). We have already remarked how the algebra of eqn.(\ref{eq:ch22-pstra}) closes only approximately while the algebra
of eqn.(\ref{eq:ch22-conftra}) closes exactly. If one had followed the strategy of PS to higher orders, one would have to continually adjust 
the actions and
transformation laws. On the contrary, our strategy, detailed in \cite{covcal2007,covcal2014}, is to keep the transformation laws \emph{fixed}
to be that of eqn.(\ref{eq:ch22-conftra}), and find a systematic calculus to generate actions invariant under it.

It is worth emphasizing the actual nature of the \emph{exact} invariances of actions like $S_{PL}$ and even the simpler $S_0^{PS}$. Denoting 
their integrands (the Lagrangean densities) by ${\cal L}_0^{PS}$ and ${\cal L}_{PL}$ it is easy to check that under, say $\epsilon^+(\tau^+)$
transformations
\be
\label{eq:ch22-confvars}
\delta_+\,{\cal L}_0^{PS}\,=\,\partial_+(\epsilon^+\,{\cal L}_0^{PS})\quad\quad\,\delta_+{\cal L}_{PL}\,=\,\partial_+(\epsilon^+\,{\cal L}_{PL})\,+\,\partial_-(\partial_+^2\epsilon^+\,{\cal L}_{PL})
\ee
In other words, only the respective actions are invariant. These are typical of general coordinate variations.

Another important aspect which has already been discussed is that the PS and PL actions are related by a \emph{field redefinition}. This
guarantees that the physical predictions of the two are the same (upto some order in $R^{-1}$), though the transformation laws take very
different forms. If in our new strategy the transformation laws are going to be fixed, no field redefinitions will be allowed any more. This
also means that the various means of simplifying the actions that were initially proposed by PS, and subsequently employed by both
Drummond and us, can no longer be used. However, it may often prove useful to resort to field redefinitions after the relevant conserved 
quantities have been obtained. Needless to say, great care must be exercised to maintain consistency.

So far, our remarks have only been with regard to the PS-type effective theories whose leading order action $S_0$ was obtained from the
Nambu-Goto action $S_{NG}$ by fixing
the so called conformal gauge of eqn.(\ref{eq:ch22-conformgauge}). The Nambu-Goto action enjoyed the much larger \emph{world-sheet
general coordinate invariance} while after fixing the conformal gauge, the \emph{residual} general coordinate transformations that leave
the conformal-gauge action invariant are precisely those of eqn.(\ref{eq:ch22-conftra}). The Nambu-Goto action could serve as the starting
point for both the L\"uscher-Weisz effective theory (in the static gauge) and the PS-effective action iin the conformal gauge.

This suggests to base the starting point for the covariant calculus on the Nambu-Goto type actions invariant under the full world-sheet
general coordinate invariance although the description does not involve any \emph{intrinsic metric} on the world sheet. Another starting
point would be the Polyakov description with an intrinsic metric for the world-sheet included, but whose invariances are extended to both
world-sheet general coordinate invariance as well as \emph{local} Weyl invariance, as already introduced in eqn.(\ref{eq:ch22-polyaact}) and
eqn.(\ref{eq:ch22-weylscaling}). We shall separately discuss both. We shall essentially follow \cite{covcal2007,covcal2014} after correcting
some errors (which we shall point out in what follows) and presenting the results in a different, logically more transparent way.
\subsection{Covariant Calculus I: The Nambu-Goto way}
\label{subsec:ch22-NGcovcal}
The crux of this approach (see \cite{covcal2007,covcal2014} for more details) is that general coordinate invariance can be implemented even
in the absence of any intrinsic world-sheet metric as long as objects can be constructed that have the same transformation property under
general coordinate transformations as an intrinsic metric. The \emph{induced metric} (already introduced in eqn.(\ref{eq:ch22-metriceq})) is
indeed one such:
\be
\label{eq:ch22-induced}
g_{\alpha\beta}\,=\,\partial_\alpha\,X^\mu\partial_\beta\,X_\mu
\ee
The reader is referred to \cite{covcal2007,covcal2014} for other important aspects of this choice. In what follows, both for the Nambu-Goto way and the Polyakov way, invariance under \emph{Target Space} Poincare transformations of the $X^\mu$
are always assumed. It should however be remembered that static gauges break this invariance manifestly. The infinitesimal forms of the
world-sheet general coordinate transformations $\xi^\alpha\rightarrow\,\xi^\alpha-\epsilon^\alpha(\xi)$ induce the transformations
\be
\label{eq:ch22-infgencov}
\delta_{gen}(\epsilon)\,X^\mu=\,\epsilon^\alpha\,\partial_\alpha\,X^\mu
\ee
satisfying the algebra
\be
\label{eq:ch22-gencovalgebra}
[\delta_{gen}(\epsilon_1),\delta_{gen}(\epsilon_2)]\,=\,(\epsilon_1^\beta\partial_\beta\epsilon_2^\alpha\,-\,\epsilon_1^\beta\partial_\beta\epsilon_1^\alpha)\partial_\alpha\,X^\mu
\ee
It is easy to see that the transformations of eqn.(\ref{eq:ch22-conftra}) are special cases of these.

Now the construction of generally covariant actions essentially follows their analogs in General Relativity. This involves the construction of
the Riemann-Christoffel symbols (see eqn.4.6.8 of Weinberg's book on Gravitation and Cosmology \cite{weingravcosmo})
\be
\label{eq:ch22-christoffeleff}
\Gamma^\delta_{\alpha\beta}\,=\,\frac{g^{\delta\kappa}}{2}\{\partial_\alpha\,g_{\kappa\beta}\,+\,\partial_\beta\,g_{\kappa\alpha}\,-\,\partial_\kappa\,g_{\alpha\beta}\}
\ee
and \emph{Covariant Derivatives}(see eqn.4.6.8 and 4.6.10 of \cite{weingravcosmo})
\be
\label{eq:ch22-covders}
V_{\beta;\alpha}\equiv\,D_\alpha\,V_\beta\,=\,\partial_\alpha\,V_\beta\,-\,\Gamma^\delta_{\alpha\beta}\,V_\delta;\quad\, V^\beta{;\alpha}\equiv\,D_\alpha\,V^\beta\,=\,\partial_\alpha\,V^\beta\,+\,\Gamma^\beta_{\alpha\delta}\,V_\delta
\ee
Generalizations to arbitrary tensors can be found in \cite{weingravcosmo}. Covariant derivatives involving only the metric tensor and the
Christoffel connection are the Riemann-Curvature tensors(see \cite{weingravcosmo}). In two dimensions, these take the particularly simple form
\be
\label{eq:ch22-riemanneff}
R_{\alpha\beta\gamma\delta}\,=\,\frac{R}{2}(g_{\alpha\gamma}\,g_{\beta\delta}\,-\,g_{\alpha\delta}\,g_{\beta\gamma})
\ee
with $R$ being the curvature scalar. Thus one need to consider $R$ and it's covariant derivatives only.

An important relation, called the metric condition
\be
\label{eq:ch22-metriccond}
g_{\alpha\beta;\gamma}\,=\,0
\ee
follows automatically from the various formulae given above.

Covariant actions can be  built out of the scalar curvature and other curvature invariants though the simplest choice
\be
\label{eq:ch22-gaussbonnet}
I_{GB}\,=\,\sqrt{-g}\,R
\ee
leads to action that is a topological invariant in two dimensions. A particularly intriguing covariant action involving only $R$ is the
Polyakov-Liouville action itself, written in a manifestly covariant form
\be
\label{eq:ch22-lioucov}
S_{Polya}\,=\,\int\,d^2\xi\,(\sqrt{-g}\,R)\,\{\frac{1}{\nabla^2}(\sqrt{-g}R)\}(\xi)
\ee
where $\nabla^2$ is the scalar-Laplacian. We shall comment on this later.

The covariant actions that can be built manifestly out of covariant derivatives of $X^\mu$ take the generic form
\be
\label{eq:ch22-covderacts}
I_{cov2}\,=\,\sqrt{-g}\,D_{\alpha_1\beta_1\ldots}X^{\mu_1}\,D_{\alpha_2\beta_2\ldots}X^{\mu_2}\dots\,D_{\alpha_n\beta_n\ldots}X^{\mu_n}\,
A^{\alpha_1\beta_1\ldots\alpha_2\beta_2\ldots}B_{\mu_1\mu_2\ldots}
\ee
with $A$ composed of two-dimensional Levi-Civita symbols as well as the metric $g_{\alpha\beta}$ and B composed of the invariant tensors
of the target space. Drummond \cite{drumm2004} had proposed the following four actions arising at order $R^{-6}$ and higher (recall that
$L\,=\,\partial_+X\cdot\partial_-X$):
\br
\label{eq:ch22-drummond4}
M_1\,&=&\,\frac{1}{L^3}\partial_+^2X\cdot\partial_+^2X\,\partial_-^2X\cdot\partial_-^2X\nonumber\\
M_2\,&=&\,\frac{1}{L^3}\partial_+^2X\cdot\partial_-^2X\,\partial_+^2X\cdot\partial_-^2X\nonumber\\
M_3\,&=&\,\frac{1}{L^4}\,\partial_+^2X\cdot\partial_-^2X\,\partial_-X\cdot\partial_+^2X\,\partial_+X\cdot\partial_-^2X\nonumber\\
M_4\,&=&\,\frac{1}{L^5}\,(\partial_-X\cdot\partial_+^2X)^2(\partial_+X\cdot\partial_-^2X)^2
\er
We illustrate the possible covariantisation of the first two of these:
\br
\label{eq:ch22-drummcov}
{\cal M}_1\,&=&\,\sqrt{g}\,D_{\alpha_1\beta_1}X\cdot\,D_{\alpha_2\beta_2}X\,D^{\alpha_1\beta_1}X\cdot\,D^{\alpha_2\beta_2}X\nonumber\\
{\cal M}_2\,&=&\,\sqrt{g}\,D_{\alpha_1\beta_1}X\cdot\,D^{\alpha_1\beta_1}X\,D_{\alpha_2\beta_2}X\cdot\,D^{\alpha_2\beta_2}X
\er
We discuss the systematic generation of higher order actions shortly.

Now these covariant actions can be worked out in various gauges of interest like the static gauge, conformal gauge etc. We shall return to
those issues after we have developed the covariant formalism for the Polyakov approach with intrinsic metric.
\subsection{Covariant Calculus II: The Polyakov Way}
\label{subsec:Polcovcal}
In the leading order of this approach the action is given by eqn.(\ref{eq:ch22-polyaact}) which is invariant under the local Weyl-scalings
of eqn.(\ref{eq:ch22-weylscaling}) over and above the invariance under world-sheet reparametrizations (general coordinate invariance). Now we
show how to extend Polyakov's considerations for effective string theories involving higher derivatives of both $X^\mu$ as well as the
intrinsic metric $h_{\alpha\beta}$. All the details can be found in \cite{covcal2007,covcal2014}.

It turns out that this extension requires considerably more powerful technical and conceptual tools than what Polyakov had to do. Since higher
derivatives of the intrinsic metric are involved, it becomes necessary to introduce covariant derivatives for local Weyl-scalings also. In
effect this amounts to new \emph{connections} which play the role of \emph{gauge fields} for local Weyl-invariance.

One starts by introducing the notion of a \emph{Weyl-scaling dimension}, also called Weyl-weight, for every field. A world-sheet tensor $\phi$  of Weyl-scaling 
dimension j transforms under local Weyl-scaling according to
\be
\label{eq:ch22-weyldim}
\phi(\xi)\rightarrow\,\phi^\prime(\xi)\,=\,\omega(\xi)^j\,\phi(\xi)
\ee
Accordingly the intrinsic metric $h_{\alpha\beta}$ has Weyl-scaling dimension of 1. Now consider, for example, a world-sheet vector $V_\beta$
with Weyl-weight $j_V$. The covariant derivative of $V_\beta$ with respect to the reparametrisations is now  given in terms of the 
$h_{\alpha\beta}$ unlike the eqn.(\ref{eq:ch22-covders}):
\be
\label{eq:ch22-covderintrinsic}
V_{\beta;\alpha}\equiv\,D_\alpha\,V_\beta\,=\,\partial_\alpha\,V_\beta\,-\,\Gamma^\delta_{\alpha\beta}\,V_\delta
\ee
with the Christoffel connection now given in terms of the intrinsic metric:
\be
\label{eq:ch22-polyachris}
\Gamma^\delta_{\alpha\beta}\,=\,\frac{h^{\delta\kappa}}{2}\,\{\partial_\alpha\,h_{\kappa\beta}\,+\,\partial_\beta\,h_{\kappa\alpha}\,-\,\partial_\kappa\,h_{\alpha\beta}\}
\ee
Now the covariant derivative, as it stands, does not have a well-defined Weyl-weight. There are two sources to this difficulty; one is that
derivatives of $V_\beta$ are involved, and the other is that derivatives of $h_{\alpha\beta}$ are also involved through $\Gamma^\delta_{\alpha\beta}$.

This suggests the introduction of a \emph{Weyl-covariant derivative} of tensors that will have the \emph{same} Weyl-weight as the tensors
themselves. From the experiences with gauge theories this is chosen to be of the form
\be
\label{eq:ch22-weylcovder}
\Delta_\alpha\,\phi\equiv\,\partial_\alpha\,\phi\,-\,j\,\chi_\alpha\,\phi
\ee
where $\chi_\alpha$ transforms like a world-sheet vector. The mathematical requirement of Weyl-covariance of $\Delta_\alpha$ can be stated 
as:
\be
\label{eq:ch22-weylcovariance}
(\Delta_\alpha\,\phi)^\prime\,=\,\omega^j\,\Delta_\alpha\,\phi
\ee
It is not difficult to see that this can be achieved provided $\chi_\alpha$ transforms under Weyl-scalings as
\be
\label{eq:ch22-weylgaugetra}
\chi_\alpha^\prime\,=\,\chi^\alpha\,+\,\partial_\alpha\,\ln{\omega}
\ee
which is indeed the way an abelian gauge-field transforms. This at once leads to a natural generalization of the Christoffel connection
of eqn.(\ref{eq:ch22-polyachris}) to something more appriate for the present context:
\be
\label{eq:ch22-weylgenconn}
G^\delta_{\alpha\beta}\,=\,\frac{h^{\delta\kappa}}{2}\,(\Delta_\alpha\,h_{\kappa\beta}\,+\,\Delta_\beta\,h_{\kappa\alpha}\,-\,\Delta_\kappa\,h_{\alpha\beta})
\ee
It will turn out to be convenient to split G according to
\be
\label{eq:ch22-weylgenconn2}
G^\delta_{\alpha\beta}\,=\,\Gamma^\delta_{\alpha\beta}\,+\,W^\delta_{\alpha\beta}\quad\quad\,W^\delta_{\alpha\beta}\,=\,\frac{1}{2}(h_{\alpha\beta}\,\chi^\delta\,-\,\delta^\delta_\alpha\,\chi_\beta\,-\,\delta^\delta_\beta\,\chi_\alpha)
\ee
Since $\Gamma^\delta_{\alpha\beta}$ transforms like a \emph{connection}, and $W^\delta_{\alpha\beta}$ transforms like a \emph{tensor}, this
split helps in establishing the important property that $G^\delta_{\alpha\beta}$ in itself transforms like a connection. From it's very
construction, it is easy to see that the Weyl-weight of $G^\delta_{\alpha\beta}$ is actually zero (in other words, it is invariant under local
Weyl-scalings) though neither $\Gamma^\delta_{\alpha\beta}$ nor $W^\delta_{\alpha\beta}$ has any well-defined Weyl-weight!

With all these preliminaries, we now give the construction of what we called \emph{Weyl-reparametrisation covariant derivatives} of, say,
a world-sheet tensor $T_{\beta_1\ldots\beta_n}$ of Weyl-weight j (generalization to arbitrary tensor types follows exactly as in General Relativity(GR)):
\be
\label{eq:ch22-weylreparcov}
{\cal D}_\alpha\,T_{\beta_1\ldots\beta_n}\,\equiv\,\Delta_\alpha\,T_{\beta_1\ldots\beta_n}\,-\,G^\kappa_{\alpha\beta_1}\,T_{\kappa\beta_2\ldots\beta_n}\,-\,\ldots\,G^\kappa_{\alpha\beta_n}\,T_{\beta_1\ldots\beta_{n-1}\kappa}
\ee
Another representation of the same object that will be useful is
\be
{\cal D}_\alpha\,T_{\beta_1\ldots\beta_n}\,=\,D_\alpha\,T_{\beta_1\ldots\beta_n}\,-\,j\,\chi_\alpha\,T_{\beta_1\ldots\beta_n}\,-\,W^\kappa_{\alpha\beta_1}\,T_{\kappa\beta_2\ldots\beta_n}\,-\ldots\,W^\kappa_{\alpha\beta_n}\,T_{\beta_1\ldots\beta_{n-1}\kappa}
\ee
In the first representation, every term has the same Weyl-weight as that of $T_{\beta_1\ldots\beta_n}$ while none of them transforms like a
tensor under reparametrisations. In the second representation, every term transforms as a tensor under reparametrisations while none of them
has a definite Weyl-weight! Putting together, one concludes that ${\cal D}T$ transforms covariantly under both reparametrisations as well as
Weyl-scalings. 

With the help of these Weyl-reparametrisation covariant derivatives one can proceed to construct actions that have Weyl-weights zero and 
that also transform as scalar densities under reparametrisations. We gather below some important properties of the variety of covariant
derivatives constructed so far. They all obey the Leibnitz rule:
\be
\label{eq:ch22-leibnitz}
D_\alpha\,T_1T_2\,=\,D_\alpha\,T_1\cdot\,T_2\,+\,T_1\cdot\,D_\alpha\,T_2 
\ee
with identical relations for $\Delta_\alpha$ and ${\cal D}_\alpha$. ${\cal D}$ also satisfies the metric condition
\be
\label{eq:ch22-calDmetric}
{\cal D}_\alpha\,h_{\beta\gamma}\,=\,0
\ee
Once again, there are two classes of manifestly covariant actions that can be constructed. The first of these is the exact analogs of the 
actions built with  the Riemann curvature tensor in the Covariant Calculus-I. The generalized curvature tensor can be constructed in
complete parallel (as for example by evaluating $[{\cal D}_\alpha,{\cal D}_\beta]$ on generic tensor fields). The explicit form is given by:
\be
\label{eq:ch22-curlDcurvature}
{\cal R}^\alpha_{\beta\gamma\delta}\,=\,\Delta_\gamma\,G^\alpha_{\beta\gamma}\,-\,\Delta_\delta\,G^\alpha_{\beta\gamma}\,+\,G^\alpha_{\gamma\eta}G^\eta_{\delta\beta}\,-\,G^\alpha_{\delta\eta}G^\eta_{\gamma\beta}
\ee
The  other is through various Weyl-reparametrisation covariant derivatives of $X^\mu$.
\subsection{Weyl Connections and Weyl-weight Compensators}
\label{subsec:ch22-connscompense}
We have so far not specified what we called the \emph{Weyl Connection}, $\chi_\alpha$, so far. As long as it transforms as a world-sheet vector
and satisfies eqn.(\ref{eq:ch22-weylgaugetra}), it's specific form was not required. The main objective of our covariant calculus is to
produce, in a systematic manner, actions that are scalar densities under reparametrisations, and are Weyl-scaling(local) invariant i.e. they
should carry zero Weyl-weights. While the former can be achieved with the help of what has been developed so far in this chapter, the latter
still needs further scrutiny.

In fact any $\chi_\alpha$ of the form
\be
\label{eq:ch22-weylconn}
\chi_\alpha\,=\,\frac{\partial_\alpha\,\ln{\Phi}}{W_\Phi}
\ee
with $\Phi$ a world-sheet scalar of Weyl-weight $W_\Phi$ would be a good candidate. Clearly, this still leaves many possibilities open for 
$(\Phi,W_{\Phi})$. To appreciate that, let us recast the above equation in a more suggestive way as ${\cal D}_\alpha\,\Phi\,=\,0$. That there
are non-trivial solutions to this follows on noting that ${\cal L}\,=\,h^{\alpha\beta}\,\partial_\alpha\,X\cdot\partial_\beta\,X$ indeed 
satisfies this with $W_{{\cal L}}\,=\,-1$. By the Leibnitz rules discussed earlier all choices of the type $\Phi_n\,=\,{\cal L}^n$
with $W_{\Phi_n}\,=\,-n$ are equally good candidates. In fact, given any two candidates $(\Phi_1,W_{\Phi_1})$ and $(\Phi_2,W_{\Phi_2})$,
it is easy to see that $(\Phi_1\Phi_2, W_{\Phi_1}+W_{\Phi_2})$ is also a candidate!

To highlight the many issues involved, let us consider a contravariant vector $V^\alpha$ with Weyl-weight J (since covariant and contravariant
tensors are mapped using $h_{\alpha\beta}$ and its inverse, their Weyl-weights are in general different), and its Weyl-reparametrisation
covariant derivative
\be
\label{eq:ch22-contracurlD}
{\cal D}_\alpha V^\beta\,=\,\nabla_\alpha\,V^\beta\,-\,J\chi_\alpha\,V^\beta\,+\,W^\beta_{\alpha\gamma}\,V^\gamma
\ee
Hence
\be
\label{eq:ch22-contracurlDiv}
{\cal D}_\alpha\,V^\alpha\,=\,\nabla_\alpha\,V^\alpha\,-\,J\chi_\alpha\,V^\alpha\,+\,W^\alpha_{\alpha\gamma}\,V^\gamma
\ee
The reparametrisation part is given by 
$\nabla_\alpha\,V^\alpha\,=\,\frac{1}{\sqrt{h}}\,\partial_\alpha(\sqrt{h}\,V^\alpha)$. In fact this is what gives the result that
$\sqrt{h}\,\nabla_\alpha\,V^\alpha$ is actually an ordinary four-divergence, a very important result in proving charge conservation in GR.
We wish to establish a similar result for $\sqrt{h}\,{\cal D}_\alpha\,V^\alpha$. On using eqn.(\ref{eq:ch22-weylgenconn2}), it follows
that $W^\gamma_{\alpha\gamma}\,=\,-\chi_\alpha$,and, on using eqn.(\ref{eq:ch22-weylconn}), one obtains
\be
\label{eq:ch22-contracurlDiv2}
{\cal D}_\alpha\,V^\alpha\,=\,\frac{\Phi^{\frac{J+1}{W_\Phi}}}{\sqrt{h}}\,\partial_\alpha(\sqrt{h}\,\Phi^{-\frac{J+1}{W_\Phi}}\,V^\alpha)
\ee
This expression brings out the central issue that scalar densities do not always come with zero Weyl-weights! Thus, $\sqrt{h}{\cal D}_\alpha\,V^\alpha$, though a scalar density under reparametrisations, has the non-zero Weyl-weight $J+1$. Now we see that if $\sqrt{h}{\cal D}_\alpha\,V^\alpha$ were also to maintain the ordinary four-divergence property, it has to be scaled(multiplied) by the exact factor of $\Phi^{-\frac{J+1}{W_\Phi}}$.

The possibility of $\Phi\,=\,{\cal L},W_\Phi\,=\,-1$ mentioned earlier is the simplest choice that one can make. It also fulfills what we
called a \emph{Denominator Principle} in \cite{covcal2007,covcal2014}. It is clear that many actions for effective string theories will
simply be unacceptable for a fundamental theory of strings as these actions will become singular for certain string configurations. But in
the effective description actions are permissible as long as they are not singular for the classical string configuration $X_{cl}$ in the
above and small fluctuations around it. In the PS-effective action the denominator was $L^2$ which clearly satisfies the denominator
principle. In the Drummond higher-order actions too all denominators were powers of L. ${\cal L}$ is the closest analog of L. We shall clarify
this more as we go along.

Returning to the four-divergence (ordinary) from $\sqrt{h}{\cal D}_\alpha\,V^\beta$, with $\Phi\,=\,{\cal L}$ as the compensator, it is
${\cal L}^{J+1}\sqrt{h}{\cal D}_\alpha\,V^\alpha$ that is an ordinary four-divergence.

To throw more light on these issues, let us construct a manifestly covariant action under Covariant Calculus-II(the Polyakov way) that is
analogous to, say, ${\cal M}_2$ of eqn.(\ref{eq:ch22-drummcov}). A first guess would be
\be
\label{eq:ch22-drum2covcal2}
{\bar{\cal N}}_1\,=\,\sqrt{h}\,\{{\cal D}_{\alpha_1\beta_1}X\cdot{\cal D}_{\alpha_2\beta_2}X\,h^{\alpha_1\alpha_2}\,h^{\beta_1\beta_2}\}^2
\ee
Though a scalar-density under reparametrisations, it's Weyl-weight is -3, making it unsuitable for a covariant action. A Weyl-weight
compensator of ${\cal L}^{-3}$ renders it into an acceptable action:
\be
\label{eq:ch22-drum2covcal2p}
{\cal N}_1\,=\,\sqrt{h}\,{\cal L}^{-3}\{{\cal D}_{\alpha_1\beta_1}X\cdot{\cal D}_{\alpha_2\beta_2}X\,h^{\alpha_1\alpha_2}\,h^{\beta_1\beta_2}\}^2
\ee
We shall see the consistency of the choice $\Phi\,=\,{\cal L}$ with the denominator principle when we examine the manifestly covariant actions
in the light cone gauge.
\section{Gauge Fixing the Covariant Actions}
\label{sec:ch22-covgaugefix}
\subsection{The Static Gauge}
\label{subsec:ch22-covstatic}
We make some brief remarks about the static gauge for the covariant actions. As is well known, quantisation requires gauges to be fixed. At
a classical level, the effective actions can be worked out in this gauge exactly as was done for the Nambu-Goto action. Even this lead to an
infinite sequence of actions the first few of which were analysed in the pioneering works of L\"uscher and Weisz 
\cite{luscherweisz2002,luscherweisz2004}. Even at the classical level we found how the underlying world-sheet reparametrisation invariance
of the Nambu-Goto action led to very important constraints between the coefficients of the L\"uscher-Weisz type of effective actions
with only the transverse fields.

The same procedure can now be followed for the actions given by the Covariant Calculus-I. Though the calculations will become increasingly
unwieldy as is typical of most higher order calculations, the procedure is unambiguous. It was tacitly assumed that the resulting effective
actions are of a \emph{local} type in a QFT-sense. But the manifestly covariant action of eqn.(\ref{eq:ch22-lioucov}) raises some reservations
in this respect. In the conformal gauge, this led to the local action of eqn.(\ref{eq:ch22-polyaliou}) which was at the heart of all the
important conclusions reached by the PS-effective string approach. The chief among them were the validity of the truncated Arvis potential
for $D\ne\,26$, and, the absence of any corrections at the level of $R^{-3}$ to this potential. Consequently, this action must play an
equally important role in the static gauge calculations too. But it is hard to see how this can produce local actions at all orders. The
extensive works by Aharony and collaborators(more on them later) in this gauge should clarify this vexing issue.
\subsection{Covariant Calculus I: The Conformal Gauge}
\label{subsec:ch22-covcalIconf}
The conformal gauge in this case, as already described before, is defined by $g_{++}\,=\,g_{--}\,=\,0$, and, $g_{+-}=g_{-+}=\partial_+X\cdot\partial_-X\,=\,L$. The inverse of the metric is given by $g^{+-}=g^{-+}=L^{-1}$. The denominator principle mentioned before can also be taken
as the requirement that the metric be non-singular.

The non-vanishing components of the Christoffel connection and the curvature tensor are easily worked out:
\br
\label{eq:ch22-conncurvI}
{\Gamma^{(1)}}^+_{++}\,&=&\,\partial_+\,\ln{L}\quad\quad {\Gamma^{(1)}}^-_{--}\,=\,\partial_-\,\ln{L}\nonumber\\
R^+_{+-+}\,&=&\,R^-_{-+-}\,=\,-R^+_{++-}\,=\,-R^-_{--+}\,=\,\partial_{+-}\,\ln{L}
\er
where we have used the superscript $(1)$ to denote that the Christoffel symbols pertain to Covariant Calculus I and are calculated with the
induced metric. The resulting scalar curvature is
\be
\label{eq:ch22-scalarcurvI}
R\,=\,-2\frac{\partial_{+-}\,\ln{L}}{L}\quad\quad\,\sqrt{g}\,R\,=\,-2\partial_{+-}\,\ln{L}
\ee
Polchinski and Strominger, in arriving at their effective action, started with the Liouville action and substituted for the conformal factor
$e^\phi$ the component $g_{+-}$ of the induced metric. This certainly appeared ad hoc. Our covariant calculus avoids making such jumps. By
substituting eqn.(\ref{eq:ch22-scalarcurvI}) in the non-local effective action where by $R$ one means the scalar curvature evaluated from
$g_{\alpha\beta}$, and on noting that the scalar Laplacian in this gauge is just $\partial_{+-}$, one straight away recovers $S_{PL}$.

We end this subsection by writing down the explicit expressions for a few non-trivial covariant derivatives of $X^\mu$. Others can be easily 
worked out using the expressions for the Christoffel connection.
\br
\label{eq:ch22-covdersI}
D_{++}X^\mu\,&=&\,\partial_{++}X^\mu\,-\,\partial_+\ln{L}\,\partial_+X^\mu\quad\quad\, D_{--}X^\mu\,=\,\partial_{--}X^\mu\,-\,\partial_-\ln{L}\,\partial_-X^\mu\nonumber\\
D_{+-}X^\mu\,&=&\,D_{-+}X^\mu\,=\,\partial_{+-}\,X^\mu
\er
\subsection{Covariant Calculus II: The Conformal Gauge}
\label{subsec:ch22-covcalIIconf}
Now we turn to a discussion of the conformal gauge for the Covariant Calculus II. The details are fascinatingly different from those for 
Covariant Calculus I! The local Weyl-rescaling allows the entire $h_{\alpha\beta}$ to be set equal to the \emph{flat} metric! We give the 
details of this conceptually very important aspect. The infinitesimal reparametrisations now act on both $(X,h)$ according to
\be
\label{eq:ch22-gencoII}
\delta_\epsilon\,X^\mu=\,\epsilon^\alpha\,\partial_\alpha\,X^\mu;\,\delta_\epsilon\,h^{\alpha\beta}\,=\,\epsilon^\gamma\,\partial_\gamma\,h^{\alpha\beta}\,-\,\partial_\gamma\epsilon^\alpha\,h^{\gamma\beta}\,-\,\partial_\gamma\,\epsilon^\beta\,h^{\alpha\gamma}
\ee
The infinitesimal Weyl-scalings do not change $X^\mu$ but act on $h_{\alpha\beta}$ according to
\be
\label{eq:ch22-weylII}
\delta_\lambda\,X^\mu\,=\,0\quad\quad\,\delta_\lambda\,h_{\alpha\beta}\,=\,\delta\,\lambda\,h_{\alpha\beta}
\ee
It is known from GR that in two dimensions, every intrinsic metric can be brought to a \emph{conformally flat} metric. The local Weyl-rescaling
can then be used to bring the intrinsic metric to a \emph{flat} metric. It is important to appreciate that both these require \emph{large}, and
not infinitesimal, transformations. Thus in the conformal gauge for Calculus-II, first one brings the intrinsic metric to 
$h_{\alpha\beta}\,=\,\eta_{\alpha\beta}$.

Introducing the coordinates $\tau^\pm\,=\,\tau\pm\,\sigma$, the non-vanishing components of the fully fixed intrinsic metric becomes
\be
\label{eq:ch22-gfintr}
h_{+-}=h_{-+}=2\quad\quad\,h^{+-}=h^{-+}=\frac{1}{2}
\ee
As can easily be verified, this does not fix the freedom to make reparametrisations and local Weyl-scalings fully. Combined infinitesimal 
reparametrisations $\epsilon_\alpha$ and infinitesimal Weyl-scalings $\delta\,\lambda$ such that
\be
\label{eq:ch22-residII}
\delta\,\lambda\,\eta_{\alpha\beta}\,=\,\partial_\alpha\epsilon_\beta\,+\,\partial_\beta\epsilon_\alpha
\ee
still preserve $h_{\alpha\beta}\,=\,\eta_{\alpha\beta}$. These \emph{residual} transformations constitute the conformal transformations in 
Calculus-II.

In this conformal gauge, with the coordinates $\tau^\pm$, ${\cal L}\,=\,\partial_+X\cdot\partial_-X\,=\,L$. Consequently,
\be
\label{eq:ch22-weylconnconf}
\chi_+,=\,-\partial_+\ln{L}\quad\quad\, \chi_-,=\,-\partial_-\ln{L}
\ee
Furthermore, the Christoffel
connections of eqn.(\ref{eq:ch22-polyachris}) now vanish identically(as they should, for flat space!):
\be
\label{eq:ch22-polyachrisfix}
{\Gamma^{(2)}}^\gamma_{\alpha\beta}\,=\,0
\ee
where we have used the superscript $(2)$ to distinguish these Christoffel connections from those of Calculus-I. The $W^\gamma_{\alpha\beta}$
in this conformal gauge are given by(there is a typo in \cite{covcal2007,covcal2014}):
\be
\label{eq:ch22-Wconf}
W^\gamma_{\alpha\beta}\,=\,-\frac{1}{2L}\,(\eta_{\alpha\beta}\partial^\gamma\,-\,\delta^\gamma_\alpha\partial_\beta\,-\,\delta^\gamma_\beta\partial_\alpha)L
\ee
It's components explicitly written down are:
\be
\label{eq:ch22-Wconfcomp}
W^+_{++}\,=\,-\chi_+\,=\,\partial_+\ln{L}\quad\quad\, W^-_{--}\,=\,-\chi_-\,=\,\partial_-\ln{L}
\ee
It is instructive to examine the structure of ${\cal D}_\alpha$ in this gauge. Consider a tensor $T_{\alpha\beta\ldots}$ of Weyl-weight j.
The components of it's Weyl-reparametrisation covariant derivatives are:
\be
\label{eq:ch22-curlDconf}
{\cal D}_+\,T^{(j)}_{\dots}\,=\,\partial_+\,T^{(j)}_{\ldots}\,-\,j\,\chi_+\,T^{(j)}_{\ldots}\,-\,t_+\,W^+_{++}T^{(j)}_{\ldots}
\ee
where $t_+$ is the number of + indices of T. Substituting the value of $W^+_{++}$ from above, this takes the rather simple form
\be
\label{eq:ch22-curlDconf2}
{\cal D}_+\,T^{(j)}_{\dots}\,=\,\partial_+\,T^{(j)}_{\ldots}\,+\,(t_+-\,j)\chi_+\,T^{(j)}_{\ldots}
\ee
and a similar expression for $+\leftrightarrow\,-$. 

Now one can work out the expressions for various covariant derivatives of $X^\mu$ and construct the required actions in this covariant gauge
for Calculus-II, but we will prove a powerful result in the next section demonstrating the complete equivalence of the conformal gauges
for both classes of the covariant calculus.
\section{Equivalence of Conformal Gauges}
\label{sec:ch22-confequiv}
We shall now demonstrate that the Weyl-reparametrisation covariant derivatives ${\cal D}_{\alpha\beta\ldots}X^\mu$ are the same in value as the
covariant derivatives $D_{\alpha\beta\ldots}X^\mu$. The equivalence is first shown for the larger class of tensors with Weyl-weight zero, of
which $X^\mu$ are the simplest examples. Let $T_{\beta_1\ldots\beta_n}$ be such a zero Weyl-weight tensor. Then,
\br
\label{eq:ch22-covderequiv}
{\cal D}_\alpha\,T_{\beta_1\ldots\beta_n}\, &=&\,\partial_\alpha\,T_{\beta_1\ldots\beta_n}\,-\,G^\gamma_{\alpha\beta_1}\,T_{\gamma\beta_2\ldots}\,-\,\ldots\nonumber\\
&=&\,\partial_\alpha\,T_{\beta_1\ldots\beta_n}\,-\,W^\gamma_{\alpha\beta_1}\,T_{\gamma\beta_2\ldots}\,-\,\ldots\nonumber\\
&=&\,\partial_\alpha\,T_{\beta_1\ldots\beta_n}\,-\,{\Gamma^{(1)}}^\gamma_{\alpha\beta_1}\,T_{\gamma\beta_2\ldots}\,-\,\ldots\nonumber\\
&=&\,D_\alpha\,T_{\beta_1\ldots\beta_n}
\er
The first step was just invoking the definition of ${\cal D}$; the second step made use of the fact that all components of $\Gamma^{(2)}$
are zero; the third step made use of the fact that all the components of $W^\gamma_{\alpha\beta}$ are equal to those of ${\Gamma^{(1)}}^\gamma_{\alpha\beta}$.

This is a striking result and has many important consequences. One of these is that all the components of ${\cal R}^\alpha_{\beta\gamma\delta}$
equal those of $R^\alpha_{\beta\gamma\delta}$. It is worth clarifying this equality in some more detail. It may superficially appear that 
${\cal R}$ defined, say, through
\be
\label{eq:ch22-calRweight}
[{\cal D}_\alpha,{\cal D}_\beta]\,V_\mu^{(j)}\,=\,-{{\cal R}^{(j)}}^\sigma_{\mu\beta\alpha}\,V_\sigma^{(j)}
\ee
may have a j-dependence. On writing $V_\mu^{(j)}\,=\,\Phi^{\frac{j}{W_\Phi}}\,V_\mu$ where $V_\mu$ is a zero Weyl-weight tensor, and on using
${\cal D}_\alpha\,\Phi\,=\,0$ as noted before, it can easily be shown that ${\cal R}$ is j-independent. Furthermore, ${\cal R}$ is also a
zero Weyl-weight tensor (these two statements have different meanings). Consequently, all the ${\cal D}$ covariant derivatives of ${\cal R}$
equal the covariant derivatives of the curvature tensor in Calculus-I. Therefore all actions constructed in the two different calculi are
the same.

This is not just an accident, and can in fact be understood in a more general way. Consider another internal metric 
$h_{+-}^\prime\,=\,\Phi\,h_{+-}$. This has the effect of making the new $\Phi^\prime\,=\,1$ and consequently $\chi_\alpha^\prime\,=\,0$. 
Therefore the transformed $W$
vanishes i.e. ${W^\prime}^\gamma_{\alpha\beta}\,=\,0$. However the transformed Christoffel connection does not vanish anymore. Instead,
${\Gamma^{(2\prime)}}^\gamma_{\alpha\beta}\,=\,{\Gamma^{(1)}}^\gamma_{\alpha\beta}$. However, the sum, $G\,=\,W\,+\,\Gamma^{(2)}$ is unchanged,
which is not surprising as G is a tensor with zero Weyl-weight.

Now one can repeat the strategy we had adopted in proving the equivalence for ${\cal D}$-derivatives of tensors with zero Weyl-weights. Even
though we proved the equivalence for Weyl-reparametrisation covariant derivatives of tensors with zero Weyl-weights, the proof can be
easily extended even to tensors with arbitrary Weyl-weights. This completes the proof of the exact equivalence of conformal gauges for
the two Covariant Calculi. It must be emphasized that these equivalences hold at purely classical levels with the understanding that the
calculus be used to construct various classical actions which are then to be quantized as was done in the earlier parts of this chapter.

As an explicit realization of this equivalence, consider eqn.(\ref{eq:ch22-drum2covcal2p}). Evaluating it in the conformal gauge for 
Calculus-II yields:
\be
\label{eq:ch22-drum2covcal2conf}
{\cal N}_1\,=\,\frac{1}{2L^3}\,(D_{++}X\cdot\,D_{--}X\,+\,D_{+-}X\cdot\,D_{+-}X)^2
\ee
Likewise, evaluating ${\cal M}_2$ of eqn.(\ref{eq:ch22-drummcov}) in the conformal gauge for Calculus-I gives:
\be
\label{eq:ch22-drum2covcal1conf}
{\cal M}_2\,=\,\frac{4}{L^3}\,(D_{++}X\cdot\,D_{--}X\,+\,D_{+-}X\cdot\,D_{+-}X)^2
\ee
Thus both represent the same action. The equality of all the covariant derivatives in the two calculi was crucial for this.
\section{Drummond Actions As Examples}
\label{sec:ch22-drummhigher}
As a way of illustrating these ideas we show how covariantisation of the order $R^{-6}$ actions proposed by Drummond \cite{drumm2004,drummreply}
works according to our Covariant Calculus-I. As clearly emphasized, in our covariant calculi the transformation laws, valid to all orders,
are held fixed. Therefore, terms proportional to the leading order equations of motion(EOM) can not be thrown away as that would be tantamount
to performing field redefinitions which would generically alter the transformation laws. However, Drummond had arrived at these candidate
forms after dropping EOM terms and any comparison can only be made upto terms proportional to the EOM's.

The Covariant Calculus-I conformal gauge expressions for ${\cal M}_1,{\cal M}_2$ after ignoring terms proportional to $D_{+-}X$ (by 
eqn.(\ref{eq:ch22-covdersI}) these are proportional to leading order EOM) are:
\br
\label{eq:ch22-drumcov1conf}
{\cal M}_1\,&=&\,\frac{2}{L^3}\{D_{++}X\cdot\,D_{++}X\,D_{--}X\cdot\,D_{--}X\,+\,(D_{++}X\cdot\,D_{--}X)^2\}\nonumber\\
{\cal M}_2\,&=&\,\frac{4}{L^3}\,(D_{++}X\cdot\,D_{--}X)^2
\er
The combination ${\cal M}_1\,-\,\frac{{\cal M}_2}{2}$ takes the form
\be
\label{eq:ch22-drumconfcomb}
{\cal M}_1\,-\,\frac{{\cal M}_2}{2}\,=\,\frac{2}{L^3}(D_{++}X\cdot\,D_{++}X)(D_{--}X\cdot\,D_{--}X)
\ee
In accordance with the proposals made by PS for constructing higher-order actions, Drummond had also dropped the leading order constraints
$\partial_\pm\,X\cdot\partial_\pm\,X$ and their derivatives. Modulo such terms the combination above is just the $M_1$ of
eqn.(\ref{eq:ch22-drummond4}). After some tedious algebra it can likewise be shown that
\be
\label{eq:ch22-drumconfcomb2}
{\cal M}_2\,=\,M_2\,-\,2M_3\,+\,M_4
\ee
More of tedious algebra shows that the covariant calculus can not produce any other combinations of $M_i$.

In summary, upto order $R^{-6}$ the most general conformal gauge action possible is given by
\be
\label{eq:ch22-conf6order}
S_6^{conf}\,=\,\int\,\frac{d^2\sigma}{4\pi}\,\{\frac{L}{a^2}\,+\,\beta\,\frac{\partial_+L\,\partial_-L}{L^2}\,+\,\beta_1(M_2\,-\,2M_3\,+\,M_4)\,+\,\beta_2\,M_1\}
\ee
It is useful to record the fully covariant version(which can be evaluated in any gauge) of this:
\be
\label{eq:ch22-cov6order}
S_6^{cov}\,=\,\int\,\frac{d^2\sigma}{4\pi}\,\{S_{NG}\,+\,\frac{\beta}{16\pi}\,(\sqrt{-g}\,R)\,\frac{1}{\nabla^2}(\sqrt{-g}R)(\xi)\,+\,\beta_2\,{\cal M}_1\,+\,(\beta_1\,-\,\frac{\beta_2}{2})\,{\cal M}_2\}
\ee

This should be considered as one of the triumphs of our Covariant Calculus. This entire action is invariant, to all orders in $R^{-1}$ under
the simple transformations of eqn.(\ref{eq:ch22-conftra}). The first term \emph{entirely} accounts for the Nambu-Goto action. This is unlike
the case in the static gauge where just the Nambu-Goto action leads to an infinite number of action terms with increasing number of 
derivatives. The second term is the equivalent of the PS-action and as already elaborated fixes $\beta$ at $\beta_c\,=\,\frac{D-26}{12}$.
Consequently, the effect of this action is to produce \emph{universal} terms to all orders, and in this case to $R^{-5}$ order. Thus, for
closed strings where one does not have to worry about any boundary term complications, the static potential and the spectrum to order $R^{-6}$
(and most likely $R^{-7}$) is determined by only two parameters $\beta_1,\beta_2$, which have no reason to be universal. These important
conclusions follow without any detailed calculations. Detailed calculations in both the static gauge and the conformal gauge have upheld these
expectations as will be discussed in the next section.
\section{Spectrum of Effective Strings at Even Higher Orders}
\label{sec:ch22-higherorders}
The natural question that arose after Drummond's, and, Hari Dass and Matlock's works that showed that the spectrum of effective strings 
upto order $R^{-3}$ was of the same form as the truncated Arvis potential upto the same order, but now with the crucial difference being
the validity in $D\ne\,26$, was whether this remarkable feature would survive even higher order corrections. Here we separately discuss the
results obtained by i) Aharony and his collaborators on the one hand, and by ii) the author and collaborators  on the other. The narratives
are not in chronological order.
\subsection{Results by Aharony et al}
\label{subsec:ch22-aharonyhigher}
Aharony and Karzbrun were the first to announce results for the spectrum of effective string theories at order $R^{-5}$ 
\cite{ahakarz2009,ahazomar2013}.
They chose to work in the static gauge that L\"uscher and Weisz had used in their pioneering work in \cite{luscherweisz2002}. They essentially
followed LW by systematically generating higher-derivative action terms, assuming \emph{locality}. Recall that LW had introduced two such
terms with coefficients $c_2,c_3$ at the level of four-derivative terms, symbolically denoted as ${\cal L}_4$. Aharony and Karzbrun first write
down the most general six-derivative terms which fall into two categories, ${\cal L}_{6,4}$ which involve four transverse fields (which they
designate as $X^i$ instead of $h^i$ as done by LW), and, ${\cal L}_{6,6}$ with 6 $X^i$. The former is of two kinds with coefficients $c_4,c_5$, of which
they argue that $c_5$ terms do not contribute. The latter has three types of terms with coefficients $c_6,c_7,c_8$ of which they show that 
$c_8$ type of action is a linear combination of $c_6$ and $c_7$ types. They go on to show, rather like the situation with the $c_2,c_3$ terms
that all but $c_4$ are contained in the Nambu-Goto action itself. They also study carefully constraints among the $c_i$ in close parallel to
the open-closed string duality of LW. Like LW, they also make extensive use of partition functions.

They too chose to work with closed strings avoiding possible complications arising from boundary terms. This is a highly detailed work 
examining strings in various backgrounds and is a treasure-house of techniques. An important fallout was that they too showed the 
\emph{absence} of $R^{-3}$ corrections to the spectrum of the Nambu-Goto action i.e. the Arvis spectrum in the static gauge also. 
Like LW, they too did not address
the question of validity of the results for $D\ne\,26$. More specifically, they did not carry out the consistency requirements as in
string theory  in the static gauge which is how the question of validity in $D\ne\,26$ has to be addressed.

They could not fix the coefficient $c_4$ which was subsequently addressed by Aharony, Field and Klinghofer \cite{ahafiekli2012} who
analysed the $R^{-5}$ corrections within the PS-formalism that made use of many insights obtained by Drummond and us. We will come to that 
shortly. There is type of non-uniformity in Aharony and Karzbrun's treatment; while they use the fully covariant Nambu-Goto action to
analyse part of the problem, they did not try to base the whole analysis on a systematic construction of fully covariant effective actions
as done by us in our Covariant Calculus. As already mentioned, they would then have to face the problem of possible non-local actions
in the static gauge. We shall return to this once more after commenting on the Polchinski-Strominger approach adopted by Aharony, Field
and Klinghofer \cite{ahafiekli2012}.

Now we comment on their work. They extend the $R^{-3}$ analysis of Drummond and us to order $R^{-5}$ within the PS-formalism, which they
funnily call the \emph{orthogonal} gauge while the correct nomenclature ought to be the \emph{conformal gauge} or even the 
\emph{covariant gauge}. This, however, does not dent the correctness of their analysis. Theirs is a more or less routine extension of the $R^{-3}$ order analysis
except for a few important technical differences. As already discussed, the correctness of the $R^{-3}$ analysis hinged on the fact that the
string momentum $P^\mu$ does not receive any corrections at this order. So these authors address the issue of higher order corrections to 
string momentum. By a clever choice of field redefinitions they show that $P^\mu$ does not get corrected at $R^{-5}$ level also.

Then they construct the Virasoro generators to this order, which now involve quartic order expressions in the oscillators. To quantize, they
adopt what they call a \emph{Weyl Ordering}. They do not seem to have explicitly verified that they satisfy the correct Virasoro algebra. Their
two most important results are: i) that the ground state energy even to this order agrees with the Arvis potential expanded to this order(two
orders higher in $R^{-1}$ than the truncated Arvis potential) but now valid for $D\ne\,26$, and, ii) the spectrum of the (1,1) excited states
show a \emph{deviation} from the Arvis spectrum to this order (except in $D=3$). 

Another important conclusion these authors reach is that their conformal gauge results to order $R^{-5}$ are in full agreement with the earlier
static gauge calculations of Aharony and Karzbrun provided their $c_4$, the only parameter they had not fixed, is taken to be 
$c_4\,=\,\frac{D\,-\,26}{192\pi}\,=\,\frac{\beta_c}{16\pi}$. This raises several interesting points and it clearly points to the
correctness of both the $R^{-5}$ order results. Firstly, it points to the origin of $c_4$ in the static gauge to be the same as that of
the PS-action, equivalently what we called the Polyakov-Liouville action (actually Aharony, Field and Klinghofer based their analysis on this
form rather than the form originally given by PS). This can only be consistent if the static gauge actions were also derived from the same
parent fully covariant action i.e. of eqn.(\ref{eq:ch22-lioucov}). But this covariant action was exactly \emph{local} only in the conformal
gauge. So one has to understand how it could have given rise to the local $c_4$ term in the static gauge. In particular, it is important to
understand at what order, if at all, the static gauge starts producing non-local actions.

It is worth emphasizing that the results of Aharony, Field and Klinghofer are completely in accordance with the expectations of our Covariant 
Calculus-I as compactly encapsulated in eqn.(\ref{eq:ch22-conf6order}) and eqn.(\ref{eq:ch22-cov6order}) \cite{covcal2007,covcal2014}. That is,
at $R^{-5}$ there are \emph{no} free parameters and there is complete universality in the sense that the spectrum only depends on D. What is
remarkable is that while the ground state energy even to this order only depends on D through the combination $(D-2)$, reflecting the 
transverse nature of the physical degrees of freedom, the excited state energies do not. As per the Covariant Calculi, \emph{potential}
non-universal terms can only arise at order $R^{-7}$ and higher, as also borne out by explicit calculations \cite{ahazomar2013}. We say
\emph{potential} because these too could, in principle, show universality. As of now, the additional parameters $\beta_1,\beta_2$ of
eqns.(\ref{eq:ch22-conf6order}, \ref{eq:ch22-cov6order}) are free parameters.  

\subsection{Alleged Equivalence to Arvis Spectrum To All Orders}
\label{subsec:ch22-ourall}
Around the same time as the work of Aharony and Karzbrun, this author, along with his collaborators Peter Matlock and Yashas Bharadwaj
had made the rather extraordinary claims that i) the spectrum to \emph{all} orders of the Polyakov-Liouville action of 
eqn.(\ref{eq:ch22-polyaliou}) was \emph{identical} to the Arvis Spectrum \cite{liouall2009}, ii) that the inclusion of the two independent 
type of Drummond actions of eqn.(\ref{eq:ch22-drumcov1conf}) still does not change the spectrum from the Arvis spectrum \cite{drummall2009}, 
and, iii) that inclusion of all the actions of Covariant Class-I also does not change the spectrum \cite{allall2009}. 

The author has himself been sceptical of these results but repeated and careful scrutiny has not revealed the fault lines. The author would like
to highlight the main lines of reasoning so that this could be resolved. More so, because the methods used were very straightforward and
mathematically elegant. It should be emphasized that the weakest link in the chain of reasoning is i) itself as ii) and iii) depend on the
correctness of i).

So let us look at the salient features of (i). The first step consists in determining the equations of motion $E^\mu$ for the $X^\mu$ arising
from this action, and in applying N\"other theorem to find the conserved $T_{--}$, to all orfers, as a result of the
exact invariance under eqn.(\ref{eq:ch22-conftra}) (likewise $T_{++}$). The reader is referred to \cite{liouall2009} for all the formulae 
and details. As already noted in our analysis of the $R^{-3}$ level corrections,
this $T_{--}$ is \emph{superficially} not \emph{holomorphic} i.e. a function of $\tau^-$ only as there are terms with $+$-derivatives of
fields. This issue is addressed by making use of eqn.(\ref{eq:ch22-generalY}) (recall that $Y^\mu$, the fluctuation field, is defined through
$X^\mu\,=\,X^\mu_{cl}\,+\,Y^\mu$) which we repeat here for ease of reading:
\be
\label{eq:ch22-generalY2}
Y^\mu(\tau^+,\tau^-)\,=\,F^\mu(\tau^+)\,+\,G^\mu(\tau^-)\,+\,H^\mu(\tau^+,\tau^-)
\ee
On substituting this in $T_{--}$, one can decompose it into $T_{--}\,=\,T_{--}^h\,+\,T_{--}^{nh}$ where the \emph{non-holomorphic}
part $T_{--}^{nh}$ has \emph{no} holomorphic parts (i.e. functions of $\tau^-$ only), though it can have anti-holomorphic parts (i.e. 
functions of $\tau^+$ only). The $T_{--}$(and likewise $T_{++}$) satisfies $\partial_+\,T_{--}\,=\,-2\pi\,E\cdot\,\partial_-X$; in
other words, $T_{--}$ is conserved \emph{on-shell} i.e. when the EOM $E^\mu\,=\,0$ is satisfied. Hence for on-shell, 
$\partial_+\,T_{--}^{nh}=0$. This can be solved uniquely as $T_{--}^{nh}$ has no holomorphic parts (unlike a similar eqn. for $T_{--}$ which
can not be so solved). In effect, this amounts to simply setting $T_{--}^{nh}\,=\,0$ for on-shell $T_{--}$. All this can be checked carefully
and there are no issues at this stage.

The next step involves a mode expansion of $G^\mu(\tau^-)\,=\,a\,\sum\,\alpha^\mu_m\,e^{-i\,m\,\tau^-}$. This is also quite general. The
Virasoro generators $L_m$ follow in a straightforward way (though they are lengthy expressions). A curious feature of the $L_m$'s is that
except for the terms coming from the quadratic part of the action $S_0$, every oscillator appears contracted with $e_+^\mu$. Therefore,
apart from a normal ordering of the free parts, there are no factor ordering issues. This feature was also seen in the results of 
Aharony, Field, and Klinghofer, so there was really no need for any Weyl-ordering.

The next step was determining the quantum algebra of $\alpha^\mu_m$. Here we deviated from the standard path of deriving them from the
Heisenberg commutation relations between $Y^\mu$ and their \emph{canonical} momenta. The reasons for this were many but we cite one of
the difficulties encountered; even at $R^{-2}$ level, we saw an inconsistency as $\Delta[p^\mu,p^\nu]\,+\,\partial_{\sigma\sigma}\Delta[Y^\mu,Y^\nu]\,+\,\partial_\tau\Delta[Y^\mu,p^\nu]$ did not vanish (here $\Delta[A,B]$ is the deviation from the free field values). We ascribed this
to the fact that the action was really of the \emph{higher derivative} type for which no canonical formalism exists. Instead, one would have
to use the famous Ostrogradsky formalism in its quantum version, further complicated by the presence of local invariances(see the summary
in \cite{ndhqchs2010}).

Instead, we opted for bit of a trial and error approach to guess the oscillator algebra. The first observation was that the free oscillator
algebra $[\alpha_m^\mu,\alpha_n^\nu]\,=\,m\,\eta^{\mu\nu}\,\delta_{m,-n}$ would not reproduce the Virasoro algebra. After some tedious work
we did find an oscillator algebra that would correctly reproduce the Virasoro algebra with central charge $D\,+\,12\beta$. Rather remarkably,
one could find a redefinition of $G^\mu(\tau^-)$, equivalently a highly non-linear redefinition of the oscillator strengths,  that reproduced
the free oscillator algebra. The same oscillator redefinition also brought the $L_0,{\bar L}_0$ to their form that would obtain from $S_0$.
This meant that the spectrum was identical to that of the Nambu-Goto theory.

Two lacunae about this all order proof are a) the string momentum was not analyzed to all orders, unless the field redefinitions adopted
render all corrections vanish. This indeed happened with the field redefinitions used by \cite{ahafiekli2012}, and, b) the field redefinitions
used are somehow not legitimate. As argued in \cite{fielddef2006}, field redefinitions must respect what we had called {\it X-uniformity}. 

Next, we come to the all order results for the Drummond type actions. Again one derives rather straightforwardly the EOM $E^\mu$ and the
stress tensor $T_{--}$ to all orders. It is noticed that every term in $T_{--}$ for both of the Drummond type actions involves $D_{++}X$. This is shown to
be non-holomorphic with the result that the entire $T_{--}$ for Drummond type actions is non-holomorphic and does not contribute to the 
Virasoro generators. Because of this, no further attention was paid to the oscillator algebras. This part of the proof is straightforward and is unlikely to be wrong. The meaning of such actions which do not
contribute to the Virasoro generators is something that needs to be explored further. Here too the string momentum was not analysed and the
source of error could be there. 

Finally, we come to the alleged claim of no corrections from any of the higher order terms arising out of the Covariant Calculus. While
the proof is not as straightforward as in the case of Drummond type terms, a proof was given that the $T_{--}$ from all of them are 
strictly non-holomorphic. Because of the rather detailed nature of these arguments, errors could well have been committed though that
possibility is highly unlikely. 
It is important to stress that neither in ii) nor iii) field redefinitions and oscillator algebras played any role. But in both of them the 
string momentum issues were not examined.

Now we summarize the results of \cite{ahakarz2009,ahafiekli2012} and comment on their implications for the all order results. They found
that even to order $R^{-5}$ the ground state energy was the same as that given by the Arvis formula, but the energy of the first excited
differed from that of the NG-theory, even though the deviations were still universal(see next section). This shows that the claims of 
\cite{liouall2009} about ground state energies is correct to this order but not the claim about excitation energies. This is in spite of
explicit corrections to string momentum found in \cite{ahafiekli2012}. This points to the oscillator algebra proposed in \cite{liouall2009}
to be problematic. The question remains whether the ground state energy continues to be that of Arvis even beyond this order. In \cite{ahazomar2013} it is only anticipated that corrections at $R^{-7}$ order and beyond are non-universal, but no calculations have been carried out. Also,
in calculations to this order and beyond, the universal contributions coming from the Polyakov-Liouville action have to be carefully
disentangled before the claims of \cite{liouall2009} can be properly assessed. Later, in \S 11 of this chapter, we shall discuss the path-integral calculations of \cite{ambeff2014} where, at least in the 
saddle point approximation, it appears as if the claims of \cite{liouall2009} for the ground state are correct. In the same work, they also
considered the Polyakov extrinsic curvature action \cite{polyaextrinsic1986} as the leading correction. They found, again within the saddle
point approximation, that the Arvis ground state energy was not altered. This is again in conformity with the analysis in \cite{allall2009}.

Thus there are aspects of the all order claims that seem correct, while some others, like excited state energies, that are clearly wrong.
Also, at some stage effective string theories must necessarily correct the tachyonic instability of the Arvis result. This is because QCD
is a perfectly unitary theory. A reassessment of the all order claims is clearly in order.
\section{Other Important Issues}
\label{sec:ch22-otherissues}
\subsection{ The Excited States}
\label{eq:ch22-excstates}
As in the case of the quantization of strings \cite{ggrtqstr1973}, it is important to investigate the spectrum of the excited states,
in addition to the ground state energies. Even before the quantisation of strings in \cite{ggrtqstr1973}, the framework for analysing spectrum
was already there in the Dual Resonance model as explained in detail in Chapter 16. The essential aspects of a spectrum are the 
energies(masses) and their degeneracies. Group theoretical techniques developed in the Dual Resonance models were more or less reemployed in
analysing the string spectrum too.

We first briefly review the theoretical status. Many of the formulae have already appeared earlier but have been collected here again for
ease of presentation. In the context of flux tubes, the relevant analysis was performed by Arvis \cite{arvis1983} 
though the energies of excited states was only
implicit in his work. Making it explicit is just to replace the expression for the Arvis potential $V_{arvis}(R)$ by
\be
\label{eq:ch22-arvisexc2}
E_{n,arvis}(R)\,=\,\sqrt{\{\sigma^2\,R^2\,+\,2\pi\sigma[-\frac{(D-2)}{24}\,+\,n]\}}
\ee
For later use, we truncate this also to order $\frac{1}{R^3}$ and rearrange the terms to write it as
\be
\label{eq:ch22-truncarvisexc2}
E_{n,arvis}^{trunc}(R)\,=\,V_{arvis}^{trunc}\,+\,\frac{\pi}{R}\,n\,+\,\frac{\pi^2}{R^3}\cdot\frac{1}{2\sigma}\,n(\frac{(D-2)}{12}-n)
\ee
Therefore, in the leading order (i.e. $R^{-1}$) one has uniform splitting of $\frac{\pi}{R}$ between excited states. The states at level n
are all degenerate to all orders in $R^{-1}$. 

But as already discussed this was valid only for $D=26$ and the PS-effective string approach gave the same results at order $R^{-1}$ but
valid in all D. Their analysis was only restricted to $R^{-1}$ order and they found agreement with the Arvis expression but now valid for all
D.

The next thorough analysis of the spectrum of effective string theories was undertaken by L\"uscher and Weisz in their 2004 work
\cite{luscherweisz2004}. Their work was in the static gauge. They gave a very detailed analysis of the spectrum employing powerful group 
theoretical techniques and those
of partition functions. While in $D=3$ they found agreement for the Arvis spectrum truncated to $R^{-3}$, in D higher than 3 they
obtained the general result for $n\le\,3$ (see their eqn.(6.1)):
\be
\label{eq:ch22-lwdim2spect2p}
\Delta\,E_{n,i}\,=\,\frac{\pi^2}{R^3}\,\{n[\frac{(D-2)}{12}-n]c_2\,+\,\nu_{n,i}(c_3\,+\,2\,c_2)\}
\ee
As already discussed they found one linear constraint between their $c_2$ and $c_3$ on the basis of their open-closed string duality,
but this still left one free parameter. However, on the basis of a classical analysis of the Nambu-Goto action, we have already shown how
both $c_2$ and $c_3$ get determined and on substitution of those values the above formula of LW coincides with that of truncated Arvis spectrum.
The lack of degeneracy found by them disappears whenever $c_3+2\,c_2$ vanishes, as indeed happens for the values coming from the Nambu-Goto
action.

The issue really hinged on whether there were any potential action terms at order $R^{-3}$ over and above that of the Nambu-Goto term.
This was settled by Drummond and us using the Polchinski-Strominger conformal gauge calculations. The upshot was that at even $R^{-3}$
order the spectrum agreed with the truncated Arvis spectrum in all dimensions.

Finally, in a remarkable series of works Aharony and collaborators carried the analysis to order $R^{-5}$ in both the static gauge and the 
conformal gauge. For the (1,1) excited states they reported a correction to the Arvis spectrum, now truncated at $R^{-5}$ level,of the
form
\be
\label{eq:ch22-ahaexc}
\Delta\,E_{1,1;i}\,=\,-\frac{\pi^3\,(D\,-\,26)}{48\sigma^2\,R^5}\,C_{1,i}
\ee
where $C_{1,i}$ are coefficients that vanish in $D=3$. They are also tabulated in Table 3 of Bastian Brandt's paper \cite{brandt2021} which
we will come to while reviewing the numerical results for the spectrum. A few comments are in order about this result. It is still a universal
correction in that it only depends on D but through the combination $(D-26)$ in contrast to the Arvis spectrum which only involves
$(D-2)$, the number of transverse degrees of freedom, to all orders. It also breaks the degeneracy of the Arvis spectrum.

Now we turn to some remarks about the numerical simulation results for the spectrum. One of the earliest numerical studies was by Juge, Kuti,
and Morningstar in 2003 \cite{kutiexc2003}. They had then reported deviations from the Nambu-Goto(Arvis) spectrum. The accuracies in those
days could not have probed the $R^{-5}$ levels where deviations are expected on theoretical grounds. Most likely these deviations were
systematic errors.

Bastian Brandt has very carefully investigated the spectrum but only in $D=3$ where the $R^{-5}$ corrections are absent \cite{brandt2021}. 
He has used large
Wilson loops as well as Polyakov loop correlators in his study as well as powerful techniques to control various sources of contamination
and systematic effects. He concludes that while there is qualitative agreement between the numerical results and the Arvis spectrum, there
are some quantitative deviations, particularly at large values of R. But large values of R are also numerically very demanding. If these
deviations survive further scrutiny and reduction of systematic errors, there will certainly be reasons to revisit the effective string
theories themselves.
\subsection{AdS-CFT approaches}
\label{subsec:ch22-adscft}
The so called AdS-CFT correspondence, also called the Gauge-Gravity duality \cite{vyaslargeD2013,vyasgaugegr2019}, is one of the most remarkable of connections ever made. There 
is an ever increasing literature on this and it is way beyond the scope of this book to even attempt a heuristic discussion. We refer the 
reader to the course on this subject by Francesco Benini \cite{beniniads2022} for a good introduction, which also has references to 
important sources. We shall be content with making some broad comments in so far as it directly concerns one of the essentials of our book, 
namely, the static potential between quarks and anti-quarks.

Broadly speaking, this correspondence in a rather precise sense is between gravity theories in $d+1$ dimensions and Conformal Field Theories
(CFT) on the d-dimensional boundary. More specifically, the gravity theories considered are on Anti-de Sitter(AdS) backgrounds. Benini gives
a simple motivation that is particularly relevant to our book. As we have already discussed, string theories can be made sensible even for
$D\le\,26$(the so called sub-critical theories) at the cost of introducing an additional degree of freedom i.e. the liouville field. This
can be thought of as providing an additional `dimension' $z$ and consider a metric on the enlarged space of the form
\be
\label{eq:ch22-preads}
ds^2\,=\,w(z)^2(dx_D^2\,+\,dz^2)
\ee
Now if one demands scale invariance wrt to the original coordinates i.e. $x\rightarrow\,\lambda\,x$, it can be realized provided the new
coordinate is likewise scaled i.e. $z\rightarrow\,\lambda\,z$ and the scale factor(often called the \emph{warp factor}) is chosen to be
$w(z)\,=\,\frac{R}{z}$ so that
\be
\label{eq:ch22-liouads}
ds^2\,=\,R^2\,\frac{dx_D^2\,+\,dz^2}{z^2}
\ee
which is an AdS metric.

What is of interest in the context of our book is the application of the correspondence to Wilson loops $W({\cal C})$ defined around a loop 
${\cal C}$. More precisely, it is to find the gravity analog for the gauge theory observable. The prescription is to take Nambu-Goto action
for AdS and find the minimal area surface that is bounded by ${\cal C}$. When only the classical AdS geometry is used, it should lead to
the linearly rising confining potential. The idea was then put forward to take into account various quantum fluctuations to probe corrections
to $\sigma\,R$ term; in particular the L\"uscher term. Naik \cite{naikads1999} showed that for $D=3$ the L\"uscher term with the correct
sign could be reproduced by only taking into account fluctuations in the radial AdS coordinate. Greensite and Olesen \cite{greenole1999}
 considered world-sheet fluctuations in a general supergravity background and found the L\"uscher term with the wrong sign casting doubts
on the utility of supergravity calculations for the observed L\"uscher term in lattice simulations of QCD (see the detailed discussion of this
in Chapter 21).

We refer the reader to these works for details. Additionally, F\"orste, Ghoshal, and Thiesen \cite{forsteads1999} also investigated the 
implications of the AdS-CFT correspondence for Wilson loops. While all these works were able to reproduce the L\"uscher term, they have
nowhere reached the levels the effective string theory descriptions have reached. In fact, even the absence of corrections at $R^{-2}$ level
have not been reproduced by the AdS-CFT approaches. Like in all approaches, higher order calculations in AdS-CFT become increasingly unwieldy
both technically as well as conceptually. But these must be done to complement what has been learnt from effective string theories so far. It
is also not clear how the AdS-CFT approach will account for the non-universal corrections at $R^{-7}$ order.
\subsection{Thickness of Flux Tubes}
\label{eq:ch22-thicktubes}
One of the striking aspects of the simulations by Bali et al \cite{balietal1995,balireport2001} and Haymaker et al \cite{haymakeretal1996}, 
discussed in 
Chapter 21 is that the flux tubes appear to have noticeable \emph{thickness} (see also \cite{casselleu1thick2016,cassellefinTthick2012}). This is clear from both the action and energy profiles of 
Bali et al. A precise interpretation of these broadened profiles is not so straightforward. From an effective string point of view it could
arise as both due to the fluctuations of a very narrow string and/or an intrinsic thickness to the effective string. It is also not clear
how to  disentangle these two aspects.

The topic of thickness of flux tubes has been of interest for a very long time. Already in 1981 L\"uscher, M\"unster, and Weisz had investigated
this in \cite{luscherthick1981} and had argued for a logarthmic increase in thickness with separation. Such an increase can not be
ascribed to any intrinsic thickness. Petrov and Ryutin \cite{petrovthick2015} have approached this through analysis of the high energy
scattering of protons, and find results different from \cite{luscherthick1981}.

The broadening of QCD flux tubes in $D=3$ has also been addressed from an AdS-CFT correspondence by Greensite and Olesen \cite{greenthick2000}.
Polchinski and Susskind \cite{polsussthick2001} made the intriguing suggestion that four-dimensional projection of certain \emph{thin} $AdS_5$ 
strings behave like thick strings. Vikram Vyas \cite{vyasthick2010} has further elaborated this point of view. In particular, he has
stressed the importance of \emph{massive} modes in this context. How to include such massive modes in effective string descriptions is far 
from clear. From one point of view, integrating out such massive modes would again give effective string theories of the kind we have
already considered. A resolution may be in examining observables over and above those of the spectrum. Needless to say, this important issue
requires further studies. Another idea that may turn out to be fruitful in this regard is that of \emph{Fat} strings in AdS. For this 
interesting perspective see the article by Sumit Das \cite{sumitfat2022}. The bulk of our book has dealt with the theme of going from
hadronic strings to fundamental strings; in \cite{sumitfat2022} Sumit Das explores how the journey can happen the other way too i.e. from
fundamental strings to QCD strings.

It is also very important to carry out the spectrum studies in $D=4$ where the $R^{-5}$ correction first shows up. While the SU(3) case was 
already very challenging even at the level of $R^{-3}$(see \cite{ndhpushan2005,ndhpushan2006}), going to $R^{-5}$ level may just be too
difficult. But a study of SU(2) or even $Z_2$ may already throw important light on these issues.

\section{Path Integral Quantization of Subcritical Strings}
\label{sec:ch22-subcritpath}
We now discuss approaches based on the path integral quantization of the Polyakov action of eqn.(\ref{eq:ch22-polyaact}) and possible additions
to it. We shall mainly be focussing on the works of Durhuus, Nielsen, Olesen, and Petersen \cite{dnopdual1982} on dual models in the saddle 
point
approximation, of Durhuus, Olesen, and Petersen \cite{dopstat1984} on the static potential in the Polyakov theory, and finally, 
of Ambjorn, Makeenko, and Sedrakyan \cite{ambeff2014} on effective QCD strings in the Polyakov approach. Though chronologically the first two
papers appeared well before the developments elaborated in the earlier sections of this chapter, we are discussing them the last in order to
contrast them with the later developments. Though the L\"uscher-Weisz approaches were also based on path-integral methods, they were done in
the static gauge for essentially the Nambu-Goto like theories.

Polyakov, in his pioneering work \cite{polyasubcr1981}, had stated that at a classical level, his action is completely equivalent to the
Nambu-Goto action. This could be seen on noting that the saddle point solution, which also happens to be exact, simply equates the intrinsic
metric to the induced metric(modulo an irrelevant local scale factor). But at the quantum level, there are dramatic differences. Polyakov
had himself demonstrated one of these in \cite{polyasubcr1981}; that was that his theory could be formulated consistently in dimensions
other than $D=26$. In particular, the appearance of the Liouville action for the so called sub-critical dimensions i.e. $D<26$.

The works to be discussed now \cite{ambeff2014,dnopdual1982,dopstat1984} all follow the path integral approach to quantization, as was also
done by Polyakov in \cite{polyasubcr1981}. A substantial improvement over Polyakov's original work was the incorporation of boundaries. At
the level of actions, open strings need boundary terms while close strings do not. But at the level of the world sheets, both of them have
to be described by manifolds with boundary. For open strings this manifold can be taken as the one with disk topology with rectangular
boundaries, while for closed strings the manifolds are of cylinder topology. Manifolds with boundaries bring in a number of subtleties, and
the behaviour of the Liouville field at the boundaries is essential for the saddle point solutions to exist. All these works, at one point 
or the other, invoke large $|D|$, saddle point techniques, and the so called mean field solutions(which are also based on large $|D|$). All
of them express the hope, without sufficient justifications, that even the large $|D|$ solutions will somehow be relevant even for real-life 
cases of $D=4,3$. This certainly can not be the case when comparisons are made to high accuracy numerical simulations, where even differences
between $D=3$ and $D=4$ are clearly noticeable.

In \cite{dnopdual1982}, the authors study the path-integral formulation of Polyakov's theory in the saddle point approximation. They further
restrict their analysis to the conformal gauge for the intrinsic metric. They only carry out the $X^\mu$-integrations. They find that a
saddle point solution can exist in the $D\rightarrow\,-\infty$ limit, and that too provided the intrinsic metric is singular on the boundary.
Proceeding in accordance with Polyakov's proposal they are able to construct off-shell Green functions for the scattering of spin-0 mesons.
They neglect the Faddev-Popov determinant(which gives rise to the contribution proportional to $-26$ in the Liouville action) as they are
only dealing with the $D\rightarrow\,-\infty$ case. They also find that when the external masses obey a certain condition, the functional
integration over the Liouville field completely decouples. Subsequently they find that the on-shell S-matrix elements reproduce the
Veneziano amplitudes. They offer no explanations for why a $D\rightarrow\,-\infty$ calculation should do so, nor any indications of how
corrections to the saddle point solution are to be treated systematically. 

In \cite{dopstat1984}, the authors study the static potential, of central interest to this book, in the Polyakov theory. They too resort to
saddle point techniques in the $D\rightarrow\,-\infty$ limit.  In particular, they study large Wilson loops. The boundary of the manifold 
is mapped to the Wilson loop.  In addition to the Polyakov action, they consider(for renormalization reasons) the action
\be
\label{eq:ch22-lioumu}
\mu_0^2\int_D\,d^2z\,\sqrt{g}
\ee
where D is the world sheet domain.  It should be noted that they use the notation $h_{ab}$ for the induced metric and $g_{ab}$ for the 
intrinsic metric! This term, which is not Weyl-invariant, leads to a term of the type $\mu_0^2\,e^\phi$ in the Liouville action. In our
considerations of effective string theories, such a term has not been considered. In addition, they also include a boundary action.

The authors invoke what they call a ``double saddle-point" approximation; the first of these realizes the $R\rightarrow\,\infty$ limit,
where $R$ is the quark-antiquark separation, and the second realizes the $D\rightarrow\,-\infty$ limit. They too perform the calculations
in the conformal gauge for the intrinsic metric and the saddle point is for the combined $X^\mu$ and Liouville fields. The functional integrals
are first carried out with the Liouville field fixed both in the interrior and on the boundary(where it takes the value $\psi(z)$). This leads
to the Liouville action(proportional to $D-26$). The saddle point approximations are made while carrying out the functional integrations over
the fields $\phi(z),\psi(z)$. Though they investigated the static potential for large $R$, they only determined the linearly rising 
and $R^{-1}$ pieces, When $\mu\ne\,0$ they reached the surprising conclusion that the $R^{-1}$ term actually vanishes(actually, they could not 
evaluate the saddle point exactly, and had to make some approximations)! It is important to understand the meaning of this result, if correct,
in the light of our earlier discussions of the L\"uscher term. For $\mu=0$ they obtain the $R^{-1}$ correction to the linearly-rising part
of the potential to be $-\frac{\pi\,D}{24R}$ whereas the universal L\"uscher term is $-\frac{(D-2)\pi}{24R}$. But as the authors have clearly
clarified, their large D approach can only capture the part proportional to $D$. This clearly points to the need for understanding, in a
systematic manner, corrections to their large D saddle point approximations. Large D techniques and $\frac{1}{D}$ corrections to them were
pioneered by Alvarez \cite{alvarezlargeD1981,alvarezsubD1983}. 

Now we come to the work of Ambjorn, Makeenko, and Sedrakyan \cite{ambeff2014} which addresses effective QCD strings beyond Nambu-Goto action.
They too choose to use the path-integral quantization of Polyakov theory for world sheets with boundaries mapped to the Wilson loops. This is
what one would have done had the strings in question been fundamental. This is in contrast to the way effective strings had been dealt with
in the ealier parts of this chapter i.e. treat them as being valid only for small fluctuations around the classical string configuration, and
treat the fluctuations quantum-mechanically in perturbation theory. 

They too work in the conformal gauge for the intrinsic metric, so that the Liouville field $\phi(z)$ is the only remaining degree of freedom
of the intrinsic metric.  Many technical features are akin to those of \cite{dopstat1984}. They treat the path-integral as if the full 
fluctuations of $X^\mu$-field are relevant(not just small fluctuations around the classical string configuration). On denoting the boundary 
value of $\phi(z)$(the Liouville field) by $\psi(z)$, they solve the EOM for $X^\mu$ with $\psi$ fixed and denote the solutions by 
$X_\mu^{\psi}(z)$. They integrate over the fluctuations of $X^\mu$ around $X_\mu^{\psi}$(being just a quadratic functional integration),
recovering the Liouville action of Polyakov. 

What is left are functional integrations over the Liouville field $\phi(z)$ in the interrior as well as $\psi(z)$ on the boundary. They 
simplify matters using the elegant Upper Half Plane (UHP) parametrization. Nevertheless, they are able to perform the functional integrationa
only in the $D\rightarrow\,\infty$ and that too using the saddle point approximation. In this limit, the so called mean-field approximation
becomes exact. These authors too make the claim, without adequate justification, that for the Polyakov formulation of the Nambu-Goto string, 
the mean field result is exact even for finite D. Till subleading(in D) corrections are actually calculated(as in \cite{alvarezsubD1983})
 and shown to vanish, one should keep the fingers crossed. Though such a possibility sounds extraordinary, it is still conceivable; the
Duistermat-Heckman theorem \cite{duistermatwkb1982} about conditions for exactness of WKB approximations is a case in point. We refer the 
reader to the many interesting details in \cite{ambeff2014}  and simply state their salient results.

They find the static potential to match exactly i.e. to all orders in $R^{-1}$, the Arvis result. As discussed extensively in \S 9 of this
chapter, this was also the claim made by \cite{liouall2009}. If the saddle point(mean field) results do not indeed get corrected, these are
strong vindications of \cite{liouall2009}(see \S 9 for detailed comparisons). Rather remarkably, this path-integral approach appears to bypass
the thorny issues of string momentum, oscillator algebras etc., inherent to the operator methods. As is well known, detailed comparisons
between path-integral and operator quantizations are notoriously hard. As discussed earlier, Aharony and collaborators found deviations from
the Nambu-Goto theory for the energies of the first excited states at order $R^{-5}$. Unfortunately, only the ground state energies were
addressed in \cite{ambeff2014}. Their calculations should be extended to the full spectrum, at least to the first excited states.

The authors attempt the simplest generalization of the Polyakov action by adding an extrinsic curvature action, first proposed by Polyakov
in \cite{polyaextrinsic1986}. The form of the extrinsic curvature action used by them is
\be
\label{eq:ch22-extrinact}
S_{extr}\,=\,\frac{1}{2\alpha}\,\int\,d^2z\,\sqrt{g}\Delta\,X\cdot\Delta\,X
\ee
(It should be recalled that in their notation $g$ is the intrinsic metric). Here $\alpha$ is a dimensionless constant and $\Delta$ is the
2d Laplace-Beltrami operator. However, the leading order corrections to the effective action proposed by Drummond as well as both our
covariant calculi, discussed earlier, are not quite of this form. In fact we had proposed two independent leading order corrections. One has 
to check whether there is any equivalence of one of those, or a linear combination of those, modulo EOM's and constraints of previous order,to
Polyakov's extrinsic curvature action in the specific form used by the authors. With the inclusion of the extrinsic curvature action, the 
saddle point analysis also gets considerably harder. The authors again resort to mean field analysis. They find that the for large $R$, the
extrinsic curvature terms do not alter the Arvis result for the ground state. As already commented in \S 9, this too is in conformity with
the claims made in \cite{drummall2009,allall2009}. Once again, it is of utmost importance to calculate the corrections to the mean field
results. However, an earlier calculation by Braaten and coworkers \cite{braatenextrinsic1987}, also using large D analysis, found an $R^{-4}$
correction. Around the same time as them Olesen and Yang \cite{olesenextrinsic1987} had also analysed the extrinsic curvature action, again in
the large $D$ limit and saddle point approximation. They specifically studied the static potential. As already emphasized, the saddle point
analysis even in the large D limit is quite difficult. Olesen and Yang made many further approximations which are not very systematic. They
reported results for the $R^{-1}$ correction in two distinct circumstances:a) non-perturbative, in the sense of not doing a perturbative
analysis for small fluctuations - they claim a deviation of the coefficient from the universal L\"uscher value. The high accuracy numerical
data discussed in chapter 21 do not seem to support such a deviation, but a reanalysis of the data with these specific non-perturbative
corrections in mind may be worthwhile,and b) perturbative analysis -- here they find the D-dependent piece of the standard L\"uscher term.
Once again, a thorough reanalysis of the extrinsic curvature actions is in order.
\section{Concluding Remarks}
\label{sec:ch22-conclusions}
In this book we have narrated, with necessary technical and conceptual elaborations, how strings made an appearance as the logical culmination
of a number of deep, fascinating and powerful attempts at understanding strong interactions. These were the ideas of S-matrix, Dispersion
Relations, Complex Angular Momenta and Regge Poles, Duality and the Dual Resonance Models. We have also explained the many conceptual problems
that arose with the string theoretic description of strong interactions, and how they were replaced by a relativistic quantum field theory
of quarks and gluons called Quantum Chromodynamics. The non-observance of quarks led to a crisis for this route to strong interactions too, but
the ingenious proposal of the dual superconductor mechanism pointed to a way out, at least in principle. 

A thin flux tube was at the heart of the matter. We have then described in detail in the last two chapters how this flux tube, established 
through impressive numerical simulations based on Lattice Gauge Theories, not just superficially resembles a string but to a very high 
degree of accuracy (more specifically, the first six terms in the large-R expansion of the static quark-antiquark potential)  
resembles a Bosonic String Theory even mathematically.

Never before in the history of physics has there been a situation where two radically different theories for the same physical phenomena have
eventually reached a conceptual fusion. This is what motivated the author to call this fantastic occurrence as \emph{Strings to Strings}!

A central theme to the analytical understanding of the flux tube has been that of effective string theories. In spirit as well as in their
mathematical implementation, these are like other effective descriptions which have been thoroughly treated in chapter 18. For strong
interactions, such effective descriptions were the celebrated Chiral Effective Field Theories. While in their case, the symmetry content
could be more easily understood from the point of view of the microscopic theory, QCD, in the case of effective string theories, even this
is rather obscure. That is, what is a microscopic understanding of the world-sheet general coordinate invariance from a QCD point of view?

Hopefully, the detailed picture of flux tubes given by effective string theories, albeit in a truncated version of QCD without dynamical
quarks, will shed light on these fundamental issues. Analytical approaches to flux tubes as pioneered by Adler \cite{adler1983} or Baker 
et al \cite{bakerfluxtube}, or the more recent ones of Akhmedov \cite{akhmedov} et al should be revived in the light of the progress made on the effective string 
theory front.

Despite half a century since the inception of QCD we still do not have first principles understanding of several issues like the pion mass,
pion-nucleon coupling constant, confinement etc. It is not enough for any understanding of confinement to just produce the linearly rising
confining part; it should also reproduce the sub-leading terms. Both numerical simulations as well as effective string theory calculations
should be pushed to probe even higher order corrections to shed more light on the non-universal terms in the static potential. It is obvious
that such non-universality has to be present in the static potential as at shorter distances it is governed more by asymptotic freedom
which is sensitive to non-universal features like the color group, quark representations etc. More specifically, a much better understanding
of the region in between the string-dominated region on the one hand, and asymptotic frredom dominated region on the other, is desirable.

Ironically, the very S-matrix program that fuelled the important developments in strong interactions itself becomes fuzzy as the asymptotic
states of QCD are not the quark and gluon states. So a revision of the tenets of S-matrix theory in the light of QCD is called for. On the
effective string theory side, the ideas and techniques developed by Simeone Hellerman and his collaborators \cite{hellerman} should be
more effectively (pun intended) integrated into the overall framework of effective string theories. The precise connections between
calculations in different gauges needs to be put on a stronger footing. We also did not elaborate the Polyakov approach based on the intrinsic
metric much.

One of the eventual hopes is to be able to explain as much of hadronic physics as possible in terms of a weakly interacting string theory.


\end{document}